\documentclass[10pt,conference]{IEEEtran}

\newif\ifconfversion

\confversionfalse

\usepackage[utf8]{inputenc}
\usepackage[T1]{fontenc}

\IEEEoverridecommandlockouts
\usepackage{tikz}
\usepackage{amsmath, mathtools}
\usepackage{amssymb}
\usepackage{amsthm}
\usepackage{thmtools}
\usepackage{thm-restate}
\usepackage{stmaryrd}
\usepackage{hyperref}
\usepackage{xcolor}

\usepackage{xspace}

\usepackage{booktabs}

\usepackage{verbatim}

\usepackage{listings}

\usepackage{caption}
\usepackage{subcaption}

\usepackage{balance}

\usepackage{paralist}

\usepackage{tikz}
    \usetikzlibrary{
        arrows,
        arrows.meta,
        backgrounds,
        calc,
        chains,
        decorations,
        decorations.pathreplacing,
        decorations.pathmorphing,
        decorations.markings,
        fit,
        matrix,
        patterns,
        positioning,
        shadows,
        shapes}
\usepackage{pgfplots}
\usepackage{pgfplotstable}
\pgfplotsset{compat=1.16}
\usepgfplotslibrary{units}

\tikzset{%
    n/.style={draw, circle, thick, inner sep=1mm, minimum width=7mm},
    l/.style={draw, rounded corners=1mm, thick, inner sep=1mm},
    ll/.style={draw, rounded corners=1mm, shape=rectangle split, rectangle split parts=2, thick, inner sep=1mm, font=\vphantom{Q}}, %
    e/.style={shorten >=1mm, shorten <=1mm, thick, ->, >=stealth},
    i/.style={initial, initial text={}},
    every initial by arrow/.style={thick, ->, >=stealth},
    blub/.style={draw, thick, rectangle, rounded corners, minimum size=5mm, fill=white},
    circ/.style={fill, circle, minimum size=2.5mm, inner sep=0pt}
}

\newcommand{\todo}[1]{{\color{red}TODO: #1}}
\renewcommand{\todo}[1]{}

\newtheorem{lemma}{Lemma}

\newcommand{\myvspace}[1]{\vspace{#1}}
\renewcommand{\myvspace}[1]{}

\begin{document}

\title{Leveraging LLVM's ScalarEvolution for\\Symbolic Data Cache Analysis}

\author{\IEEEauthorblockN{Valentin Touzeau}
\IEEEauthorblockA{
\textit{Saarland University}\\\textit{Saarland Informatics Campus}\\
Saarbrücken, Germany \\
valentin.touzeau@cs.uni-saarland.de}
\and
\IEEEauthorblockN{Jan Reineke}
\IEEEauthorblockA{
\textit{Saarland University}\\\textit{Saarland Informatics Campus}\\
Saarbrücken, Germany \\
reineke@cs.uni-saarland.de}
}

\maketitle

\begin{abstract}
While instruction cache analysis is essentially a solved problem, data cache analysis is more challenging.
In contrast to instruction fetches, the data accesses generated by a memory instruction may vary with the program's inputs and across dynamic occurrences of the same instruction in loops.

We observe that the plain control-flow graph (CFG) abstraction employed in classical cache analyses is inadequate to capture the dynamic behavior of memory instructions.
On top of plain CFGs, accurate analysis of the underlying program's cache behavior is impossible.

Thus, our first contribution is the definition of a more expressive program abstraction coined {symbolic control-flow graphs}, which can be obtained from LLVM's ScalarEvolution analysis.
To exploit this richer abstraction, our main contribution is the development of \emph{symbolic data cache analysis}, a smooth generalization of classical LRU must analysis from plain to symbolic control-flow graphs.

The experimental evaluation demonstrates that symbolic data cache analysis consistently outperforms classical LRU must analysis both in terms of accuracy and analysis runtime.
\end{abstract}

\maketitle


\newcommand{\Associativity}{\mathit{k}}
\newcommand{\BlockSize}{\mathit{BS}}
\newcommand{\NbSets}{\mathit{NS}}
\newcommand{\block}{\mathit{block}}
\newcommand{\set}{\mathit{set}}
\newcommand{\powerset}[1]{\mathcal{P}(#1)}

\newcommand{\blocks}{\mathcal{B}}
\newcommand{\ie}                {i.e.\ }
\newcommand{\eg}                {e.g.,\ }
\newcommand{\Eg}                {E.g.,\ }
\DeclarePairedDelimiter\round{\lparen}{\rparen}
\DeclarePairedDelimiter\brackets{\lbrack}{\rbrack}
\DeclarePairedDelimiter\angles{\langle}{\rangle}
\DeclarePairedDelimiter\ceil{\lceil}{\rceil}
\DeclarePairedDelimiter\floor{\lfloor}{\rfloor}

\section{Introduction}

Due to technological developments, the latency of accesses to DRAM-based main memory is much higher than the latency of arithmetic and logic computations on processor cores.
This ``memory gap'' is commonly tackled by a hierarchy of caches between the processor cores and main memory. 

In the presence of caches, the latency of a memory access may vary widely depending on the level of the memory hierarchy that is able to serve the access.
Hits to the first-level cache take just a few processor cycles, while accesses that miss in all cache levels and thus need to be served by main memory can take hundreds of cycles.

This variability is a challenge in the context of real-time systems, where it is necessary to bound a program's worst-case execution time (WCET)~\cite{Wilhelm08} to guarantee that safety-critical applications meet all of their deadlines.
For accurate WCET analysis it is thus imperative to take caches into account.
The timing variability induced by caches also introduces security challenges.
Implementations of cryptographic algorithms have been shown to be vulnerable to cache timing attacks~\cite{Bernstein05} and cache analysis~\cite{Doychev15,Wang17,Sung2018} may help to uncover such vulnerabilities or prove their absence.\looseness=-1

Cache analysis aims to statically characterize a program's cache behavior by classifying memory accesses in the program as guaranteed cache hits or misses.
One perspective on cache analysis is that it is the composition of two phases:
\begin{compactenum}
	\item A transformation of the program under analysis into a simpler program abstraction: a control-flow graph (CFG) whose edges are decorated with memory accesses.\looseness=-1
	\item An analysis of this decorated CFG that classifies accesses as ``always hit'', ``always miss'', or ``unknown''.
\end{compactenum}
For instruction cache analysis this two-phase approach works well, as CFGs accurately captures most programs' instruction fetch sequences. 
For data cache analysis, however, a plain CFG abstraction can be highly inaccurate.
Consider for example the following simple loop:
        \begin{lstlisting}[language=C, basicstyle=\small\ttfamily, numbersep=1em, xleftmargin=0em]
    for (int x = 0; x < 100; x++)
    	sum += A[x]
        \end{lstlisting}
In each iteration of the loop a different address is accessed, and so the corresponding edge in the CFG needs to be conservatively decorated with all possible addresses.
The order in which the array elements are accessed is lost and it becomes impossible to make accurate predictions about the program's cache behavior.
A program abstraction that more precisely captures a program's memory access behavior is thus needed.

Our first contribution is the definition of {symbolic control-flow graphs} in Section~\ref{sec:symboliccfg}, which is our formalization of the output of LLVM's ScalarEvolution analysis~\cite{ScalarEvolution,Absar2018}.
Symbolic CFGs accurately capture the link between loop iterations and accessed memory blocks via chains of recurrences~\cite{Bachmann1994,Engelen2001} in a manner that is amenable to static analysis.

To exploit this more expressive program abstraction our main contribution is the development of \emph{symbolic data cache analysis} in Section~\ref{sec:symbolic_analysis}, a smooth generalization of Ferdinand's classical LRU must analysis~\cite{Ferdinand1997,Ferdinand1999} from plain to symbolic control-flow graphs.
To fully realize the potential of symbolic data cache analysis we further introduce a context-sensitive analysis combining loop peeling and unrolling in Section~\ref{sec:peelingunrolling} and various implementation tricks in Section~\ref{sec:implementation}. 

The experimental evaluation on the PolyBench benchmark suite in Section~\ref{sec:experiments} demonstrates that symbolic cache analysis compares favorably to classical LRU must analysis both in terms of accuracy and analysis runtime.

\todo{mention sections for different contributions}

 \todo{Thank Canberk}

\section{Background}%
\label{sec:background}

\todo{distinguish memory references (static) and memory accesses (dynamic)}

\subsection{Caches}\label{sec:caches}

    Caches are fast but small memories that buffer parts of the large but slow main memory in order to bridge the speed gap between the processor and main memory.
    Caches consist of \emph{cache lines}, which store data at the granularity of memory blocks $b \in \blocks$.
    Memory blocks usually comprise a power-of-two number of bytes~$\BlockSize$, e.g. 64 bytes, so that the block $\block(a)$ that address $a$ maps to is determined by truncating the least significant bits of $a$, i.e., $\block(a) = \lfloor a/\BlockSize \rfloor$.
    In order to facilitate an efficient cache lookup, the cache is organized in \emph{sets} such that each memory block maps to a unique cache set~$\set(b) = b \bmod \NbSets$, where $\NbSets$ is the number of sets.
    The number of cache lines~$\Associativity$ in each cache set is called the \emph{associativity} of the cache.
    
    If an accessed block resides in the cache, the access \emph{hits} the cache.
    Upon a cache \emph{miss}, the block is loaded from the next level of the hierarchy.
     Then, another memory block has to be evicted due to the limited size of the cache.
    The block to evict is determined by the \emph{replacement policy}.
    In this paper, we assume the least-recently-used (LRU) policy, which replaces the block that has been accessed least recently.
    A memory block~$b$ hits in an LRU~cache of associativity~$\Associativity$ if $b$ has been accessed previously and less than $\Associativity$~distinct blocks in the same cache set have been accessed since the last access to~$b$.
    LRU is generally considered to be the most predictable replacement policy~\cite{Reineke2007}.\looseness=-1

    In this paper, we refer to the \emph{age} of block~$b$ as the number of distinct blocks in the same cache set that have been accessed since the last access to~$b$.
    Thus, an access to block~$b$ hits the cache if and only if its age is less than the associativity~$\Associativity$.

\myvspace{-1mm}
\subsection{Control-Flow Graphs as a Program Representation}\label{sec:cfg}

\todo{terminology: plain vs symbolic CFGs}

Control-flow graphs (CFGs) are a program representation commonly employed in compilers and static analysis tools.
A CFG is a directed graph $\mathcal{G}=(V, E, v_0)$, whose vertices~$V$ correspond to control locations in the program including the initial control location~$v_0 \in V$, and whose edges~$E$ represent the possible control flow between the graph's vertices.\looseness=-1

For the purpose of cache analysis, CFGs are used to represent the possible sequences of memory accesses generated by the underlying program.
To this end, each edge of the CFG is decorated with the set of memory addresses that may be accessed when control passes along that edge.

As defined above, CFGs over-approximate the behavior of the program they represent as they do not capture the functional semantics of the instructions.
In particular, all paths through the graph are assumed to be feasible even if, in reality, some are not. %
Also, and this is particularly problematic for data cache analysis, the CFG representation does not capture the dependence of the accessed memory addresses on the loop iterations.
We will see in Section~\ref{sec:illustrativeexample} how this may lead to gross overapproximations of the number of cache misses.
To overcome this issue, we introduce symbolic control-flow graphs %
in Section~\ref{sec:symboliccfg}.\looseness=-1

\myvspace{-1mm}
\subsection{Ferdinand's May and Must Cache Analysis}

The aim of Ferdinand's may and must cache analyses~\cite{Ferdinand1997,Ferdinand1999} is to classify memory accesses in a CFG as definite hits or definite misses.
As noted before, under LRU replacement, an access results in a cache hit if and only if the age of the accessed block is less than the cache's associativity.

Instead of computing all reachable concrete cache states, must and may analysis operates on abstract cache states, which maintain upper and lower bounds on the age of each memory block. 
Each block's bounds hold independently of the ages of other blocks.
This allows for a compact representation of large sets of concrete cache states.
For example, the abstract must cache state $\lambda b. \infty$ that maps every block to age bound~$\infty$ compactly represents all possible concrete cache states. %
As the correlation between the ages of different blocks is lost, the resulting analysis is not exact.
However, recent work~\cite{Touzeau2017,Touzeau2019,Stock2019} has shown that the loss in precision due to this abstraction is small in practice.
Our symbolic data cache analysis introduced in Section~\ref{sec:symbolic_analysis}, can be seen as a smooth generalization of Ferdinand's must analysis to symbolic control-flow graphs.\looseness=-1

\section{Illustrative Example}\label{sec:illustrativeexample}

\newcommand{\Entry}[1]{\mathit{entry}_{#1}}
\newcommand{\Backedge}[1]{\mathit{backedge}_{#1}}
\newcommand{\Assume}[2]{\mathit{assume}_{#1, #2}}

    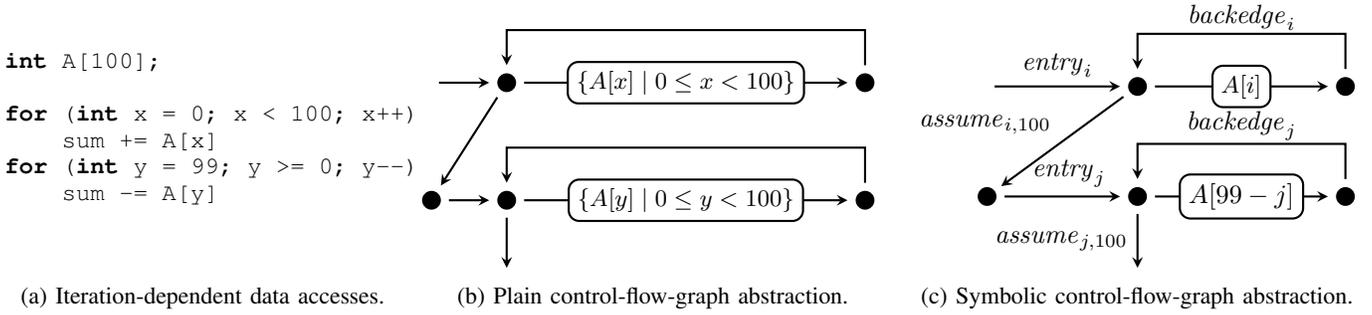
\begin{figure*}
    	\begin{subfigure}[b]{0.29\textwidth}
        \begin{lstlisting}[escapeinside={(*@}{@*)},language=C, basicstyle=\small\ttfamily, numbersep=1em, xleftmargin=0em]
int A[100];
            
for (int x = 0; x < 100; x++)
    sum += A[x]
for (int y = 99; y >= 0; y--)
    sum -= A[y]

	(*@ \hspace{0pt} @*)
        \end{lstlisting}
          \vspace{-1mm}
        \caption{Iteration-dependent data accesses.}\label{fig:motivatingprogram}
        \end{subfigure}\hfill
    	\begin{subfigure}[b]{0.37\textwidth}
		\centering
                    \begin{tikzpicture}[node distance=15mm, yscale=1]
                \node[circ] (S0) {};
                \node[circ, right=of S0, xshift=3cm] (S1) {};
                \draw[e] (S0) edge node[blub] {\small$\{A[x] \mid 0 \leq x < 100\}$} (S1);
                \draw[e] (S1) -- ($(S1)+(0, 0.7)$) -| (S0);
                \draw[e] ($(S0)+(-1, 0)$) -- (S0);

                \node[circ, below=of S0, yshift=2mm, xshift=-1cm] (S0primepre) {};
                \node[circ, below=of S0, yshift=2mm, xshift=0cm] (S0prime) {};
                \node[circ, right=of S0prime, xshift=3cm] (S1prime) {};
       
                \draw[e] (S0) to (S0primepre);
                \draw[e] (S0primepre) to (S0prime);

                \draw[e] (S0prime) edge node[blub] {\small$\{A[y] \mid 0 \leq y < 100\}$} (S1prime);
                \draw[e] (S1prime) -- ($(S1prime)+(0, 0.7)$) -| (S0prime);
    
                \draw[e] (S0prime) -- ($(S0prime)+(0, -1)$);
            \end{tikzpicture}
          \vspace{-1mm}
        \caption{Plain control-flow-graph abstraction.}\label{fig:plaincfg}
        \end{subfigure}\hfill
         \begin{subfigure}[b]{0.34\textwidth}
		\centering
                    \begin{tikzpicture}[node distance=15mm, every text node part/.style={align=center},yscale=1]
                \node[circ] (S0) {};
                \node[circ, right=of S0, xshift=1cm] (S1) {};
                \draw[e] (S0) edge node[blub] {\small$A[i]$} (S1);
                \draw[e] (S1) -- ($(S1)+(0, 0.7)$) -| (S0) node[pos=0.25,above,yshift=-.75mm]{$\Backedge{i}$};
                \draw[e] ($(S0)+(-2, 0)$) -- (S0) node[pos=0.5,above]{$\Entry{i}$};

                \node[circ, below=of S0, yshift=3mm, xshift=-2cm] (S0primepre) {};
                \node[circ, below=of S0, yshift=3mm, xshift=0cm] (S0prime) {};
                \node[circ, right=of S0prime, xshift=1cm] (S1prime) {};
   
			    \draw[e] (S0) -- (S0primepre) node[pos=0.4,left,yshift=1mm,xshift=-2mm]{$\Assume{i}{100}$};
                \draw[e] (S0primepre) -- (S0prime) node[pos=0.55,above]{$\Entry{j}$};
                    
                \draw[e] (S0prime) edge node[blub] {$A[99-j]$} (S1prime);
                \draw[e] (S1prime) -- ($(S1prime)+(0, 0.7)$) -| (S0prime) node[pos=0.25,above,yshift=-.75mm]{$\Backedge{j}$};

                \draw[e] (S0prime) -- ($(S0prime)+(0, -1.05)$) node[pos=0.5,left]{$\Assume{j}{100}$};
 
           \end{tikzpicture}
          \vspace{-1mm}
        \caption{Symbolic control-flow-graph abstraction.}\label{fig:symboliccfg}
        \end{subfigure}
        \vspace{-2mm}
        \caption{Simple program and its plain and symbolic control-flow-graph abstractions.}
    \end{figure*}

As an illustrative example of the drawbacks of cache analysis performed on plain CFGs, consider the simple program in Figure~\ref{fig:motivatingprogram}.
The first loop of our example program iterates across array $A$ in the forward direction, while the second loop iterates across the same array in the opposite direction.

\subsection{Intuitive Cache Analysis}
Let us intuitively analyze the program's cache behavior. 
For this analysis, we will assume a tiny set-associative cache with LRU replacement consisting of 2 cache sets, an associativity of 4, and cache lines of size 8 bytes.
Thus, the cache has a capacity of $2\cdot 4\cdot 8 = 64$ bytes.
Also assume that integers are of size 4 bytes, and so the cache can hold 16 array cells.\looseness=-4

Assuming an initially empty cache, the first loop does not exhibit any temporal locality, as each array cell is only touched once.
However, it does exhibit spatial locality, as pairs of adjacent array cells may reside in the same memory blocks.
Thus, every other iteration of the first loop will result in a cache hit.\looseness=-1

The second loop accesses the same array cells as the first.
Now it depends on the cache geometry whether and to what extent this temporal locality can be exploited.
Under our assumptions, the cache will contain array cells $A[84], A[85], \dots, A[99]$ after the first loop has terminated.
Thus, the first 16 iterations of the second loop hit the cache.
The remaining iterations profit only from spatial locality as the first loop did, hitting in every other iteration.

\subsection{Traditional Cache Analysis}

Under Ferdinand's must cache analysis~\cite{Ferdinand1997,Ferdinand1999} and recent exact analyses~\cite{Touzeau2017,Touzeau2019} the program is abstracted via its CFG and the CFG's edges are annotated with the sets of memory blocks that may be accessed while executing the corresponding part of the program as discussed in Section~\ref{sec:cfg}. 
Figure~\ref{fig:plaincfg} shows the plain CFG abstraction for our example program.
While this abstraction is adequate for instruction cache analysis as the same instructions are accessed in each loop iteration, it is inadequate for data cache analysis, as the link between the loop iteration and the accessed address is lost.
As a consequence, it is impossible to predict any of the memory accesses in the program to be cache hits or misses.

If the entire set of memory blocks that can potentially be accessed fits into the cache, then persistence analysis~\cite{Mueller2000rts,Cullmann2013tecs,Zhang15,Reineke18,Stock2019} may deduce that each of these blocks results in at most one cache miss.
However, in our example, the array~$A$ does not fully fit into the cache, and so persistence analysis is of no use here.

\subsection{Symbolic Control-Flow Graphs and Cache Analysis}\label{sec:illustrative_symbolic}

\paragraph{Symbolic Control-Flow Graphs}
We have seen that the plain CFG abstraction is inadequate for data cache analysis, because the link between loop iterations and accessed memory blocks is lost.
Thus, our first step towards accurate data cache analysis is to employ what we coin \emph{symbolic CFGs}, a simple yet powerful program representation that concisely captures the link between loop iterations and accessed data.
Symbolic CFGs are our formalization of the output of LLVM's ScalarEvolution Analysis~\cite{ScalarEvolution,Absar2018}. %

Figure~\ref{fig:symboliccfg} shows a symbolic CFG for our example program.
In a symbolic CFG---where possible---the addresses of memory accesses are expressed in terms of the loop iterations of their enclosing loops.
To this end, symbolic CFGs make it explicit when a loop is entered and when a new loop iteration begins.
These transitions are indicated by annotating edges with $\Entry{i}$ and $\Backedge{i}$, where $i$ is the identifier of a loop.
Consider the edge annotated with $A[99-j]$.
This is to be interpreted as follows:
In an execution of the program, let~$\sigma(j)$ be the number of times that $\Backedge{j}$ has been traversed since the last time $\Entry{j}$ has been taken.
Then, the accessed address is $A[99-\sigma(j)]$.\looseness=-1

For some loops, ScalarEvolution is also able to derive the exact number of times that a loop's back edges are taken from entry to exit.
To express such information, symbolic CFGs may contain $\Assume{i}{e}$ statements, where $e$ is an expression that may refer to loop variables other than $i$ itself.
An edge annotated with $\Assume{i}{e}$ can only be taken if the value of~$i$ is equal to the value of expression~$e$.
In our example, the back edges of both loops are taken exactly 100 times, and so the exit edges of both loops are annotated accordingly with assume statements.\looseness=-1

We define symbolic CFGs in Section~\ref{sec:symboliccfg}.
There we also discuss multivariate chains of recurrences~\cite{Bachmann1994,Engelen2001,Engelen2001b}, which are used to represent access expressions and loop bounds.

\paragraph{Symbolic Cache Analysis}
Symbolic CFGs are useful for data cache analysis as they capture a program's memory access behavior more precisely than plain CFGs.
In fact, in our example, the symbolic CFG perfectly captures the sequence of memory accesses generated by the program. 

It remains to define a static analysis that can efficiently exploit this information.
Simply applying Ferdinand's must analysis would not be fruitful as the underlying abstraction does not capture the relation between loop iterations and cache states.
A relatively straightforward approach would be to virtually unroll the loops for the sake of the analysis, resulting in an exploded plain CFG in which each edge could once more be annotated with a concrete memory access.
Ferdinand's must analysis could then be employed successfully on this exploded plain CFG.
However, this approach would be very costly, in particular for programs with large loop bounds. %
We are thus seeking a precise analysis whose runtime is independent of the loop bounds of the program.

To this end, our first basic idea are symbolic cache states that capture how cache states depend on the loop iteration.
To motivate symbolic cache states, consider Figures~\ref{sec:concstate} and \ref{sec:concstatenext}, which show the concrete cache states at the ends of iterations~$15$ and $17$ of the first loop from our example program.
As we assume cache lines of size $8$ bytes, each line contains two cells of the array.
We represent each memory block by the first array cell mapping to that block.
Our idea is to represent memory blocks \emph{symbolically} in terms of the values of loop variables.
For example, $A[14]$ can be expressed as $A[i-1]$ if $i$'s value is $15$.
If we represent the states from Figures~\ref{sec:concstate} and \ref{sec:concstatenext} in this way we arrive at the symbolic cache state depicted in Figure~\ref{sec:symbstate}.
Furthermore, the \emph{same} symbolic state will be reached at the end of each odd loop iteration, starting from iteration $15$.\looseness=-1

Like Ferdinand's must analysis our symbolic data cache analysis determines upper bounds on the ages of memory blocks.
However, instead of associating bounds with concrete memory blocks, it associates these bounds with symbolic memory blocks.
A peculiar consequence of this abstraction is that symbolic cache states also need to be updated when the value of a loop variable changes.
For example, if the back edge of the first loop is taken to move from iteration $15$ ($17, 19, \dots$) to iteration $16$ ($18, \dots)$, then the symbolic cache state needs to be updated to account for incrementing~$i$.
The resulting symbolic cache state is depicted in Figure~\ref{sec:symbstatenext}.
We show how to lift Ferdinand's analysis to symbolic cache analysis in Section~\ref{sec:symbolic_analysis}.\looseness=-1

In our example, one can observe that the symbolic cache states ``stabilize'' in odd and even loop iterations after the cache has been filled in the first $16$ iterations.
Thus the analysis needs to distinguish the first $16$ loop iterations from the rest, and odd from even loop iterations in the remainder of the execution.
This can be achieved by context-sensitive analysis~\cite{Martin1998,Mauborgne2005,Wegener2012}.
In Section~\ref{sec:peelingunrolling} we introduce a context-sensitive analysis that can be configured to virtually peel and unroll the loops appropriately for a given cache configuration.

    \begin{figure*}
    	\centering
	\begin{subfigure}[t]{0.19\textwidth}
    	\centering
        		\begin{tikzpicture}
    			\node {$\begin{tabular}{ll}
    				\toprule
    				Set 0 & Set 1\\
    				\midrule
    					$A[12]$ & $A[14]$\\
    					$A[8]$	 & $A[10]$\\
    					$A[4]$	 & $A[6]$\\
    					$A[0]$	 & $A[2]$\\
    					\bottomrule
    			    \end{tabular}$};
    		\end{tikzpicture}
		\myvspace{-1mm}
		\caption{Cache state at the end of iteration~$15$.}\label{sec:concstate}
    	\end{subfigure}
	\begin{subfigure}[t]{0.20\textwidth}
    	\centering
        		\begin{tikzpicture}
    			\node {$\begin{tabular}{ll}
    				\toprule
    				Set 0 & Set 1\\
    				\midrule
    					$A[16]$ & $A[14]$\\
    					$A[12]$	 & $A[10]$\\
    					$A[8]$	 & $A[6]$\\
    					$A[4]$	 & $A[2]$\\
    					\bottomrule
    			    \end{tabular}$};
    		\end{tikzpicture}
		\myvspace{-1mm}
		\caption{Cache state at the end of iteration~$17$.}\label{sec:concstatenext}
    	\end{subfigure}
	\begin{subfigure}[t]{0.28\textwidth}
		\centering
    		\begin{tikzpicture}
			\node {$\begin{tabular}{ll}
				\toprule
				Set $a$ & Set $b$\\
				\midrule
					$A[i-3]$ & $A[i-1]$\\
					$A[i-7]$	 & $A[i-5]$\\
					$A[i-11]$	 & $A[i-9]$\\
					$A[i-15]$	 & $A[i-13]$\\
					\bottomrule
			    \end{tabular}$};
		\end{tikzpicture}
		\myvspace{-1mm}
		\caption{Symbolic cache state at the end of iterations~$15, 17, 19, \dots$}\label{sec:symbstate}
    	\end{subfigure}
 	\begin{subfigure}[t]{0.28\textwidth}
		\centering
    		\begin{tikzpicture}
			\node {$\begin{tabular}{ll}
				\toprule
				Set $a$ & Set $b$\\
				\midrule
					$A[i-4]$ & $A[i-2]$\\
					$A[i-8]$	 & $A[i-6]$\\
					$A[i-12]$	 & $A[i-10]$\\
					$A[i-16]$	 & $A[i-14]$\\
					\bottomrule
			    \end{tabular}$};
		\end{tikzpicture}
		\myvspace{-1mm}
		\caption{Symbolic cache state at the start of iterations~$16, 18, 20, \dots$}\label{sec:symbstatenext}
    	\end{subfigure}
  	\myvspace{-1mm} 
    	\caption{Cache states that arise during the execution of the first loop.}
	\myvspace{-2mm}
    \end{figure*}
    
\newcommand{\MaxPeel}{\mathit{MaxPeel}}
\newcommand{\MaxUnroll}{\mathit{MaxUnroll}}
\newcommand{\peel}{\mathit{peel}}
\newcommand{\unroll}{\mathit{unroll}}

\section{Symbolic Control-Flow Graphs}\label{sec:symboliccfg}

We have seen the intuition behind symbolic control-flow graphs in Section~\ref{sec:illustrative_symbolic}.
One aspect that has been left undefined there is the shape of expressions used to represent memory accesses and loop bounds.
We fill this gap in Section~\ref{sec:mcrs}, which is then used in the formal definition of symbolic control-flow graphs in Section~\ref{sec:symbolic_cfg}.
In Section~\ref{sec:cachesemantics} we provide a semantics for symbolic CFGs, which will allow us to make formal correctness statements about the symbolic data cache analysis introduced in Section~\ref{sec:symbolic_analysis}.

\subsection{Multivariate Chains of Recurrences}\label{sec:mcrs}

\newcommand{\LoopVars}{\mathit{LoopVar}}
\newcommand{\eval}[2]{\llbracket #1 \rrbracket_{#2}}
\newcommand{\MCR}[1]{\mathit{M}(#1)}
\newcommand{\vars}{\mathit{vars}}
\newcommand{\init}[2]{\mathit{Init}(#1, #2)}
\newcommand{\shift}[2]{\mathit{Sh}(#1, #2)}
\newcommand{\subst}[3]{\mathit{Sub}(#1, #2, #3)}

We employ \emph{multivariate chains of recurrences}~\cite{Bachmann1994,Engelen2001,Engelen2001b} (short: MCRs) as the formalism for expressions.
Given a subset of a program's loop variables $S \subseteq \LoopVars$, the set $\MCR{S}$ of MCRs over $S$ is given by the following grammar:\looseness=-1
	\begin{equation*}
		\begin{split}
			e  := {} & n \in \mathbb{Z} \\
			   |\; & e_1 \mathop{bop} e_2 \text{ where } \mathop{bop} \in \{+,-,\cdot\}, \text{and } e_1, e_2 \in \MCR{S}\\
			   |\;  & \{e_1, +, e_2\}_i \text{ where } 
					i \in S, e_1 \in \MCR{S \setminus \{i\}}, e_2 \in \MCR{S}
		\end{split}
	\end{equation*}
Thus, expressions can (i) be constants; (ii) they can be formed from subexpressions via addition, subtraction, and multiplication; and (iii) they can be \emph{add recurrences} of the form $\{e_1, +, e_2\}_i$. 

Given an environment $\sigma : \LoopVars \rightarrow \mathbb{N}$ assigning loop variables to their values, an MCR can be evaluated as follows:
	\begin{equation*}
		\begin{split}
			\eval{n}{\sigma} &:= n \\
			\eval{e_1~\mathit{bop}~e_2}{\sigma} &:= \eval{e_1}{\sigma}~\mathit{bop}~\eval{e_2}{\sigma} \\
			\eval{\{e_1, +, e_2\}_i}{\sigma} &:=
				 \eval{e_1}{\sigma} + \displaystyle\sum_{k = 0}^{\sigma(i) - 1} \eval{e_2}{\sigma[i \mapsto k]} \\
		\end{split}
	\end{equation*}
By $\sigma[i \mapsto v]$ we denote the function that maps $i$ to $v$ and otherwise is the same as $\sigma$. 
Thus, in an add recurrence $e_1$ can be seen as the initial value, and $e_2$ as the increment.
For example:\looseness=-1
\begin{align*}
	\eval{\{23, +, 4\}_i}{\sigma} & = \eval{23}{\sigma} + \sum_{k=0}^{\sigma(i)-1} \eval{4}{\sigma[i \mapsto k]}
	=  23 + 4\cdot \sigma(i)
\end{align*}
Thus, the array access $A[i]$ from our example can be expressed as $\{A, +, 4\}_i$, assuming $A \in \mathbb{N}$ is the base address of the array and each element of the array is of size $4$.
Similarly, the array access $A[99-j]$ can be expressed as $\{A+396, +, -4\}_j$.

Nested add recurrences can represent arbitrary polynomial functions, e.g. $\eval{\{0, +, \{5, +, 1\}_i\}_j\}}{\sigma} = 5\cdot \sigma(j)+\sigma(i)\cdot \sigma(j)$ and $\eval{\{0, +, \{0, +, 2\}_i\}_i\}}{\sigma} = (\sigma(i)-1)\cdot \sigma(i)$.

In order to update symbolic cache states upon incrementing loop variables, we need a shift operation on MCRs that adapts an expression to account for the increment of a variable.
Such an operation should thus satisfy the following equality: $\eval{\shift{e}{i}}{\sigma[i \mapsto \sigma(i)+1]} = \eval{e}{\sigma}$.
For example, $\shift{\{A, +, 1\}_i}{i}$ could be $\{A-1, +, 1\}_i$. %

To implement such a shift operation we need an initialization operation that satisfies $\eval{\init{e}{i}}{\sigma} = \eval{e}{\sigma[i \mapsto 0]}$, which can be implemented as follows: 
\begin{equation*}
\begin{split}
	\init{n}{i}                      &:= n\\
	\init{e_1\ \mathit{bop}\ e_2}{i} &:= \init{e_1}{i}\ \mathit{bop}\ \init{e_2}{i}\\
	\init{\{e_1, +, e_2\}_j}{i} & :=\hspace{-1mm}\begin{cases}
							e_1	&: i = j\\
							 \{\init{e_1}{i}, +, \init{e_2}{i}\}_j\hspace{-3mm}~	&: i \neq j
						\end{cases}\\
\end{split}
\end{equation*}
This allows us to implement the $\shift{e}{i}$ operation:
\begin{equation*}
\begin{split}
	\shift{n}{i}                       &:= n\\
	\shift{e_1\ \mathit{bop}\ e_2}{i}  &:= \shift{e_1}{i}\ \mathit{bop}\ \shift{e_2}{i}\\
	\shift{\{e_1, +, e_2\}_i}{i} &:= \{e_1 - \init{\shift{e_2}{i}}{i}, +, \shift{e_2}{i}\}_i\\
	\shift{\{e_1, +, e_2\}_j}{i}         &:= \{\shift{e_1}{i}, +, \shift{e_2}{i}\}_j
\end{split}
\end{equation*}
The correctness of the $\init{e}{i}$ and $\shift{e}{i}$ can be shown by structural induction, which we omit here for brevity.

To take into account loop bounds when exiting a loop provided by assume statements in our symbolic control-flow graphs, we rely on a substitution operation with the following semantics: $\eval{\subst{e}{i}{\mathit{expr}}}{\sigma} = \eval{e}{\sigma[i \mapsto \eval{\mathit{expr}}{\sigma}]}$.
A heuristic implementation of $\subst{e}{i}{\mathit{expr}}$, which may fail on some inputs, can be realized as follows:
\begin{equation*}
	\begin{multlined}
	\subst{e}{i}{\mathit{expr}} :=\\
	\begin{cases}
		e
			& \text{if } e \in \mathbb{Z} \\
		s_1 \mathop{bop} s_2
			& \text{if } e = e_1 \mathop{bop} e_2 \wedge s_1 \neq \mathit{fail} \wedge s_2 \neq \mathit{fail}  \\
		\{s_1, +, s_2\}_j
			& \text{if }
				e = \{e_1, +, e_2\}_j \wedge j \neq i \\ &~~~\wedge~s_1 \neq \mathit{fail} \wedge s_2 \neq \mathit{fail}\\
		e_1 + e_2 \cdot \mathit{expr}
			& \text{if } e = \{e_1, +, e_2\}_i \wedge i \notin e_2\\
		\mathit{fail}
			& \text{otherwise}
	\end{cases}
	\end{multlined}
\end{equation*}
where $s_1 = \subst{e_1}{i}{\mathit{expr}}$ and $s_2 = \subst{e_2}{i}{\mathit{expr}}$.

Engelen~\cite{Engelen2001,Engelen2001b} provides a set of rewrite rules for MCRs that are proven to be confluent and terminating.
We rely on these rewrite rules to bring MCRs into a normal form.\looseness=-1

\newcommand{\unknownAccess}{\ensuremath{\textbf{X}}\xspace}

In general, not all accesses generated in a program can be accurately captured by an MCR. 
As an example, consider the accesses generated by a loop traversing a dynamic heap data structures, such as a linked list.
To soundly represent such accesses we introduce \emph{unknown accesses}, denoted \unknownAccess, which are interpreted to take any possible value.
Fortunately, \unknownAccess is only rarely needed in the analysis of real-time applications, in which dynamic data structures are uncommon. \todo{evidence for this statement?}

\subsection{Symbolic Control-Flow Graphs}\label{sec:symbolic_cfg}

\newcommand{\Decorations}{\mathcal{D}}
\newcommand{\Accesses}{\mathcal{A}}
\newcommand{\Statements}{\mathcal{S}}

A \emph{symbolic CFG} is a tuple $\mathcal{G} = (V, E, \LoopVars, v_0)$, where~$V$ is a set of vertices and $E \subseteq V \times \Decorations \times V$ is a set of edges, $\LoopVars$ is a set of loop variables, and $v_0 \in V$ is a vertex with no incoming edges marking the program entry.

Edges are decorated with \emph{accesses}~$\Accesses$ and \emph{statements}~$\Statements$, i.e., $\Decorations = \Statements \cup \Accesses$:
\begin{compactitem}
	\item Accesses are MCRs or unknowns:\\ $\Accesses := \MCR{\LoopVars} \cup \{\unknownAccess\}$
	\item Statements either mark the entry to a loop ($\Entry{i}$), a back edge of a loop ($\Backedge{i}$), or an assumption on the value of a loop variable ($\Assume{i}{e}$): %
	\vspace{-1mm}
		\begin{equation*}
			\begin{multlined}
				\Statements := \{\Entry{i}, \Backedge{i} \mid i \in \LoopVars\} \\ \cup \{\Assume{i}{e} \mid i \in \LoopVars, e \in \MCR{\LoopVars \setminus \{i\}}\}
			\end{multlined}
		\end{equation*}
\end{compactitem}

\subsection{Semantics of Symbolic Control-Flow Graphs}\label{sec:cachesemantics}

\newcommand{\pstates}{\Sigma_p}
\newcommand{\cstates}{\Sigma_c}
\newcommand{\pstate}{\sigma_p}
\newcommand{\cstate}{\sigma_c}
\newcommand{\updates}{\mathit{update}_\Statements}
\newcommand{\updatelru}{\mathit{update}_\mathit{LRU}}
\newcommand{\update}{\mathit{update}}
\newcommand{\unreachable}{{\bot_p}}

The state of an execution of a symbolic control-flow graph consists of two parts: %
The program state $\pstate \in \pstates$ and %
 the cache state $\cstate \in \cstates$.

We represent the program state by a map $\pstate$ that maps loop variables to their values.
Each loop variable counts the number of times that the loop back edge has been taken since last entering the loop.
The program semantics of a symbolic CFG is then captured by a transformer $\updates$ that captures the effects of statements on program states.
\begin{equation*}
	\begin{multlined}
		\updates(\pstate, s) := \\
		\begin{cases}
		\pstate[i \mapsto 0]
			& \text{if } s = \Entry{i} \\
		\pstate[i \mapsto \pstate(i) + 1]\hspace{-7mm}~
			& \hspace{4mm} \text{if } s = \Backedge{i} \\
		\unreachable
			& \text{if } s = \Assume{i}{\mathit{expr}} \wedge \pstate(i) \neq \eval{\mathit{expr}}{\pstate}\\
		\pstate
			& \text{if } s = \Assume{i}{\mathit{expr}} \wedge \pstate(i) = \eval{\mathit{expr}}{\pstate} \\
		\end{cases}
	\end{multlined}
\end{equation*}
Note that we use the special value~$\unreachable$ to represent unreachable program states, i.e.\ those not satisfying an assume statement.\looseness=-1

We represent cache states as maps $\cstate$ from memory blocks to ages, i.e.\ $\cstate$ tracks the age of each memory block in its cache set: $\cstate \in \blocks \rightarrow \mathbb{N}$.
The LRU replacement policy is then captured by the following transformer:
\begin{equation*}
	\begin{multlined}
		\updatelru(\cstate, b) := \lambda b' \in \blocks.\hspace{1cm} \\
		\hspace{1cm}\begin{cases}
		0
			& \text{if } b = b' \\
		\cstate(b')
			& \text{else if } \set(b) \neq \set(b')\\
		\cstate(b')
			& \text{else if } \cstate(b) \leq \cstate(b')\\
		\cstate(b') + 1
			& \text{otherwise}\\
		\end{cases}
	\end{multlined}
\end{equation*}
To paraphrase the above definition: (i) The accessed block~$b$ attains age~$0$.
(ii) The ages of blocks in other cache sets ($\set(b) \neq \set(b')$) do not change.
(iii) If the accessed block~$b$ is younger than block $b'$, then $b$ has already been accounted for in the age of $b'$, and thus the age of $b'$ should not increase.
(iv) Otherwise, $b$ maps to the same cache set as $b'$ and is older than $b'$ and thus the access increases the age of $b'$.

\newcommand{\drop}{\mathit{drop}_\unreachable}

The complete state of the system is a pair $(\pstate, \cstate)$ and we can capture its evolution upon arbitrary CFG decorations by combining the previous transformers into a single one and accounting for unknown accesses:
\begin{equation*}
	\begin{multlined}
		\update((\pstate, \cstate), d) :=\\
		\hspace{2mm}\begin{cases}
		\{(\updates(\pstate, d), \cstate)\}
			& \text{if } d \in \Statements\\
		\{(\pstate, \updatelru(\cstate, \block(\eval{d}{\pstate})))\} & \text{if } d \in \Accesses \setminus \{\unknownAccess\}\\ %
		\{(\pstate, \updatelru(\cstate, b)) \mid b \in \blocks\} & \text{if } d = \unknownAccess
		\end{cases}
	\end{multlined}
\end{equation*}
where $\block$ maps addresses to the corresponding memory blocks (see Section~\ref{sec:caches}). %
Note that $\update((\pstate, \cstate), d)$ maps to sets of states to capture the non-determinism introduced by unknown accesses.
We lift $\update$ to sets of states as follows:
 \begin{multline*}
 	\update(S, d) := \{(\pstate', \cstate') \mid (\pstate, \cstate) \in S \\\wedge (\pstate', \cstate') \in \update((\pstate, \cstate), d) \wedge \pstate' \neq \unreachable\}
\end{multline*}
We drop unreachable states (where $\pstate' = \unreachable$) here.

We define the set of reachable states at each control location~$R^C : V \rightarrow \powerset{\pstates \times \cstates}$ as the least solution to the following set of equations:
\begin{align}
	 R^C(v_0) &= \{(\lambda i. 0, \cstate) \mid \cstate \in \cstates\}\label{eqn:collecting_semantics_init}\\
	 	 \forall v \in V \setminus \{v_0\}: R^C(v) &= \bigcup_{(u, d, v) \in E} \update(R^C(u), d)\label{eqn:collecting_semantics_fixpoint}  
\end{align}

Equation~\eqref{eqn:collecting_semantics_init} captures that initially all loop variables are zero, while the initial cache state can be arbitrary.
Equation~\eqref{eqn:collecting_semantics_fixpoint} captures that the reachable states at node~$v$ are determined by the reachable states at $v'$s predecessor nodes~$u$ updated according to the CFG decoration between $u$ and $v$.
In keeping with abstract interpretation literature~\cite{Schmidt98}, we refer to $R^C$ as the \emph{collecting semantics}.

\section{Symbolic Data Cache Analysis}\label{sec:symbolic_analysis}

Explicitly computing the collecting semantics $R^C$ would be very costly and only possible at all if all loops were bounded.
In this section, we lift Ferdinand's must analysis to symbolic control-flow graphs to obtain a tractable analysis. 

\newcommand{\SymCaches}{\widehat{\mathit{SymCache}}}
\newcommand{\SymCache}{\widehat{\sigma}}
\newcommand{\dom}{\textit{dom}}

\subsection{Abstract Domain}

As described earlier, Ferdinand's must analysis maps memory blocks to an upper bound on their maximum age in order to classify memory accesses as hits.
Our analysis relies on a similar map, except that it maps symbolic blocks, represented via MCRs, to such age bounds.
Our abstract domain is thus \[\SymCache \in \SymCaches = \MCR{\LoopVars} \hookrightarrow \{0, \dots, \Associativity-1, \infty\},\]
where $\hookrightarrow$ indicates that symbolic cache states are partial functions.
We refer to the domain of a cache state~$\SymCache$, i.e., the set of MCRs for which $\SymCache$ provides an age bound, as $\dom(\SymCache)$.

If our analysis maps an MCR~$e$ to age~$x$ at program point~$v$, it means that the memory block containing the address given by $\eval{e}{\pstate}$ has age at most~$x$ for any program state $\pstate$ reachable at $v$.
This set of program and cache states associated with an abstract state~$\SymCache$ is captured by the concretization function $\gamma$:
\begin{equation}\label{eq:concretization}
	\gamma(\SymCache) := \{(\pstate, \cstate) \mid \forall e \in \dom(\SymCache): \cstate(\block(\eval{e}{\pstate})) \leq \SymCache(e)\}
\end{equation}

Similarly to the definition of the collecting semantics (see Equations~\eqref{eqn:collecting_semantics_init} and~\eqref{eqn:collecting_semantics_fixpoint}), which uses set unions to capture all possible behaviors of the program, we need a join operator on the abstract domain to summarize states from several incoming CFG edges.
This join operator~$\sqcup$ conservatively keeps, for each MCR, the maximum of the two upper bounds provided by the joined states: $\SymCache_1 \sqcup \SymCache_2 = \lambda e \in \dom(\SymCache_1)\cap\dom(\SymCache_2). \max\{\SymCache_1(e), \SymCache_2(e)\}$. 
This join operator is correct with respect to the concretization function:\looseness=-1
\begin{restatable}[Join Correctness]{lemma}{JoinCorrectness}\label{lem:JoinCorrectness}
	For all $\SymCache_1, \SymCache_2 \in \SymCaches$:
	\begin{equation*}
		\gamma(\SymCache_1) \cup \gamma(\SymCache_2) \subseteq \gamma(\SymCache_1 \sqcup \SymCache_2) 
	\end{equation*}
\end{restatable}
The proofs of all lemmas and theorems can be found in the 
\ifconfversion
extended version~\cite{arXivVersion}.
\else
appendix.
\fi

\subsection{Abstract Transformers}\label{sec:abstrans}

\newcommand{\aupdatea}{\widehat{\mathit{update}_{\Accesses \setminus \{\unknownAccess\}}}}
\newcommand{\aupdatex}{\widehat{\mathit{update}_\unknownAccess}}
\newcommand{\aupdates}{\widehat{\mathit{update}_\Statements}}
\newcommand{\aupdate}{\widehat{\mathit{update}}}

\newcommand{\alias}{\mathit{alias}}
\newcommand{\AliasRelation}{\mathit{AliasRelation}}
\newcommand{\SameSet}{\mathit{ss}}
\newcommand{\SameBlockWhenSameSet}{\mathit{sb}\textit{+}\mathit{ds}}
\newcommand{\DifferentBlock}{\mathit{db}}
\newcommand{\SameBlock}{\mathit{sb}}
\newcommand{\SameSetDifferentBlock}{\mathit{ssdb}}
\newcommand{\DifferentSet}{\mathit{ds}}

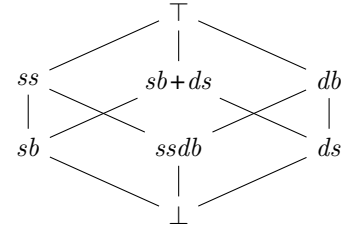
\begin{figure}
	\center
	\begin{tikzpicture}[yscale=0.9]
		\node (top)    at (2, 3) {$\top$};
		\node (ss)     at (0, 2) {$\SameSet$};
		\node (dsorsb) at (2, 2) {$\SameBlockWhenSameSet$}; %
		\node (db)     at (4, 2) {$\DifferentBlock$};
		\node (sb)     at (0, 1) {$\SameBlock$};
		\node (ssdb)   at (2, 1) {$\SameSetDifferentBlock$};
		\node (ds)     at (4, 1) {$\DifferentSet$};
		\node (bot)    at (2, 0) {$\bot$};

		\draw (top)    -- (ss);
		\draw (top)    -- (dsorsb);
		\draw (top)    -- (db);
		\draw (ss)     -- (sb);
		\draw (ss)     -- (ssdb);
		\draw (dsorsb) -- (ds);
		\draw (dsorsb) -- (sb);
		\draw (db)     -- (ds);
		\draw (db)     -- (ssdb);
		\draw (sb)     -- (bot);
		\draw (ssdb)   -- (bot);
		\draw (ds)     -- (bot);
	\end{tikzpicture}
	\myvspace{-2mm}
	\caption{Lattice of alias relations.}\label{fig:alias-relation}
	\myvspace{-3mm}
\end{figure}

To reflect the cache updates upon memory accesses, we provide two abstract transformers: $\aupdatea$, for accesses to MCRs, and $\aupdatex$, for unknown accesses.

Unknown accesses can potentially increase the age of any block in the cache.
Thus:
\begin{equation*}
	\aupdatex(\SymCache) := 
		   \lambda e' \in \dom(\SymCache). \begin{cases}
		\SymCache(e') + 1
			& \text{ if } \SymCache(e')+1 < \Associativity \\
		\infty
			& \text{ otherwise}
	\end{cases}
\end{equation*}

It is easy to prove that this transformer is correct:
\begin{restatable}[Unknown Access Transformer Correctness]{lemma}{UnknownAccessTransformerCorrectness}
	\label{lem:unknown-access-transformer-correctness}
	For all $\SymCache \in \SymCaches$, we have:
	\normalfont
	\begin{equation*}
		\update(\gamma(\SymCache), \unknownAccess) \subseteq \gamma(\aupdatex(\SymCache))
	\end{equation*}
\end{restatable}

The $\aupdatea$ transformer is similar to the one used by Ferdinand's must analysis; it rejuvenates the accessed symbolic block, and increases the ages of blocks in the same cache set that are younger than the accessed block.

The main difference lies in the fact that contrary to concrete memory blocks, which have a fixed address, it is not always obvious whether two symbolic blocks map to the same cache set or even to the same block.
We thus rely on an auxiliary function $\alias$, which, given two symbolic blocks, determines their alias relation.

\todo{MCRs vs symbolic blocks; note somewhere that we use the two terms interchangeably}

There are six possible alias relations between two MCRs:
\begin{compactenum}
	\item ``Same block'' $\SameBlock$: they map to the same memory block.
	\item ``Same set'' $\SameSet$: they map to the same cache set. %
	\item ``Different set'' $\DifferentSet$: they map to different cache sets.
	\item ``Different block'' $\DifferentBlock$: they map to different  blocks. %
	\item ``Same set, diff. block'' $\SameSetDifferentBlock:$ conjunction of $\SameSet$ and $\DifferentBlock$.
	\item ``Same block or different set'' $\SameBlockWhenSameSet$: disjunction of $\DifferentSet$ and $\SameBlock$; can also be seen as the complement of $\SameSetDifferentBlock$. %
\end{compactenum}
As shown in \cite{Hahn2011}, these relations form a lattice, whose Hasse diagram is shown in Figure~\ref{fig:alias-relation}.
The alias relation of two MCRs $e_1$ and $e_2$ can be determined as follows, where $\BlockSize$ is the size of memory blocks (in bytes) and $\NbSets$ is the number of cache sets:\looseness=-1
\begin{equation*}
	\begin{multlined}
	\alias(e_1, e_2) := \\\begin{cases}
		\SameBlock
			& \text{if } e_1 - e_2 = n \in \mathbb{Z} \wedge n = 0\\
		\DifferentSet
			& \text{else if } e_1 - e_2 = n \in \mathbb{Z}~\wedge \\
				& ~~\BlockSize \leq n \bmod (\NbSets \cdot \BlockSize) \leq (\BlockSize \cdot \NbSets) - \BlockSize\\
		\SameBlockWhenSameSet
			& \text{else if } e_1 - e_2 = n \in \mathbb{Z}~\wedge  -\BlockSize < n < \BlockSize\\
		\top
			& \text{otherwise} 
		\end{cases}
	\end{multlined}
\end{equation*}
We assume a modulo operation based on \emph{floored division}, i.e., $a \bmod n := a - n\cdot \lfloor a/n \rfloor$, so that $0 \leq a \bmod n < n$ for $n > 0$.

The alias relation between $e_1$ and $e_2$ is determined by computing the difference $n$ of the two expressions.
If the difference between $e_1$ and $e_2$ is not a constant expression, then no relation is established (last case).
Otherwise, different relations can be deduced depending on the value of $n$:
\begin{enumerate}[(i)]
	\item If $n$ is $0$, we can deduce $\SameBlock$.
	\item Addresses whose difference is a multiple of the way size ($\NbSets \cdot \BlockSize$) are guaranteed to be in the same cache set.
		Conversely, if the difference between $e_1$ and $e_2$ is more than $\BlockSize$ ``away'' from being a multiple of the way size, then $e_1$ and $e_2$ must map to different sets.
	\item If $e_1$ and $e_2$ are close, i.e., less than a block size apart, they either map to the same block or to different sets.\looseness=-1
\end{enumerate}
Other aliasing relations, such as $\SameSetDifferentBlock$ and $\DifferentBlock$ could also be deduced, but are not useful in the following.

Using $\alias$ to deduce the relation between symbolic blocks, we can formally define the transformer $\aupdatea$ to apply when performing the memory access associated with MCR~$e$.%
\begin{equation*}
\begin{multlined}
	\aupdatea(\SymCache, e) := \lambda e' \in \dom(\SymCache) \cup \{e\}.\hspace{2cm} \\
		   \hspace{2cm}\begin{cases}
		0
			& \text{ if } \alias(e, e') \sqsubseteq \SameBlock\\
		\SymCache(e')
			& \text{ else if } \alias(e, e') \sqsubseteq \SameBlockWhenSameSet\\
		\SymCache(e')
			& \text{ else if } \SymCache(e) \leq \SymCache(e')\\
		\SymCache(e') + 1
			& \text{ else if } \SymCache(e')+1 < \Associativity \\
		\infty
			& \text{ otherwise}
	\end{cases}
\end{multlined}
\end{equation*}

Unsurprisingly, the transformer closely resembles the definition of its concrete counterpart $\updatelru$.
(i) As in the concrete case, the accessed symbolic block is rejuvenated to age~0, as are all symbolic blocks that represent the same block.
(ii) A symbolic block that is in the $\SameBlockWhenSameSet$ relation to the accessed block retains its age, which is safe, as seen by the following case distinction:
Either the block is actually the accessed block and it should get age~$0$, or it maps to a different set and its age should be unchanged (first two cases of $\updatelru$).
(iii) If the accessed symbolic block~$e$ is younger than symbolic block~$e'$, then $e$ has already been accounted for in the age of $e'$, and thus the age of $e'$ should not increase.
(iv) The age of a block cannot increase by more than one upon a single access, so the fourth case is always safe.
(v) We do not distinguish ages beyond $\Associativity$, as it is not helpful to classify accesses as hits or misses. Instead we summarize these with the safe upper bound $\infty$.

As for the join operator and for unknown accesses, we prove that the access transformer is correct:
\begin{restatable}[MCR Access Transformer Correctness]{lemma}{AccessTransformerCorrectness}
	\label{lem:access-transformer-correctness}
	For all $\SymCache \in \SymCaches$ and $e \in \Accesses$, we have:
	\normalfont
	\begin{equation*}
		\update(\gamma(\SymCache), e) \subseteq \gamma(\aupdatea(\SymCache, e))
	\end{equation*}
\end{restatable}

The $\aupdatea$ transformer described above captures the effect of memory accesses.
As the symbolic cache states are tied to the program state via the concretization function given in (\ref{eq:concretization}), changes to the loop variables need to be accounted for by appropriately adapting our symbolic cache states.
We thus provide a second transformer, $\aupdates$, which captures the effect of program statements on symbolic cache states.

We define $\aupdates$ separately for each type of statement. %
The case of a back edge is arguably the most interesting one.
Each symbolic block $e$ needs to be replaced by its shifted version when $i$ is incremented, so that the expression preserves its original value%
, which is achieved as follows:\looseness=-1
\begin{equation}\label{eq:aupdatebackedge}
	\begin{multlined}
		\aupdates(\SymCache, \Backedge{i}) := \{(\shift{e}{i}, b) \mid (e,b) \in \SymCache\}
	\end{multlined}
\end{equation}
For example, $\shift{\{A, +, 4\}_i}{i} = \{A-4, +, 4\}_i$, which corresponds to replacing $A[i]$ by $A[i-1]$ upon incrementing~$i$.
One might wonder whether the set defined in Equation~(\ref{eq:aupdatebackedge}) actually defines a function.
This is indeed the case for MCRs in normal form~\cite{Engelen2001,Engelen2001b} for which $\shift{\cdot}{i}$ is bijective.

Entering a loop entails resetting the corresponding loop variable to $i$.
However, unless the prior value of $i$ is known, there is no way of rewriting expressions involving the variable~$i$ accordingly. Thus, in such cases the information for the corresponding MCRs is discarded:
\begin{equation*}
	\begin{multlined}
		\aupdates(\SymCache, \Entry{i}) := \{(e, b) \mid (e, b) \in \SymCache \wedge i \not\in e\}
	\end{multlined}
\end{equation*}

\newcommand{\reduce}{\mathit{red}}

Finally, assume statements allow the analysis to substitute the corresponding loop variable by the assumed expression.
This allows to retain information across multiple loops or in nested loops, e.g. in our running example where data cached in the first loop is reused in the second loop.
\begin{equation*}
\begin{multlined}
	\aupdates(\SymCache, \Assume{i}{\mathit{expr}}) := \\ %
								 \reduce(\{(e', b) \mid (e, b) \in \SymCache \wedge e' = \subst{e}{i}{\mathit{expr}} \neq \mathit{fail}\}),
\end{multlined}
\end{equation*}
where $\reduce(S) := \{(e, b) \mid (e,b) \in S \wedge \forall (e,b') \in S: b' \geq b\}$.

The substitution may result in multiple expressions becoming equal, e.g., $\subst{\{0, +, 2\}_i}{i}{10} = \subst{\{10, +, 1\}_i}{i}{10}$.
Then $\reduce(S)$ keeps the best bound and thereby ensures that the resulting relation is still a function.\looseness=-1

This abstract transformer for statements is also correct:
\begin{restatable}[Statement Transformer Correctness]{lemma}{StatementTransformerCorrectness}
	\label{lem:statement-transformer-correctness}
	For all $\SymCache \in \SymCaches$ and $s \in \Statements$, we have:
	\begin{equation*}
		\update(\gamma(\SymCache), s) \subseteq \gamma(\aupdates(\SymCache, s))
	\end{equation*}
\end{restatable}

\subsection{Analysis Correctness and Termination}

We can now merge the statement and access transformers into a single one that deals with the three kinds of decorations:
\begin{equation*}
	\begin{multlined}
		\aupdate(\SymCache, d) :=
		\begin{cases}
			\aupdates(\SymCache, d)
				& \text{if } d \in \Statements \\
			\aupdatea(\SymCache, d)
				& \text{if } d \in \Accesses \setminus \{\unknownAccess\} \\
			\aupdatex(\SymCache, d)
				& \text{if } d = \unknownAccess
		\end{cases}
	\end{multlined}
\end{equation*}

Similarly to the collecting semantics %
we define the abstract semantics as the least solution %
of the following equations:
\begin{align}
	\widehat{R}(v_0) &= \emptyset \label{eqn:abstract_semantics_init} \\
	\forall v \in V \setminus \{v_0\}: \widehat{R}(v) &= \bigsqcup_{(u, d, v) \in E} \aupdate(\widehat{R}(u), d) \label{eqn:abstract_semantics_fixpoint}
\end{align}

Equations~\eqref{eqn:abstract_semantics_init} and \eqref{eqn:abstract_semantics_fixpoint} are the abstract counterpart of Equations~\eqref{eqn:collecting_semantics_init} and~\eqref{eqn:collecting_semantics_fixpoint}.
We can now state the main correctness theorem about our analyzer, which follows by standard Abstract Interpretation arguments from Lemmas~\ref{lem:JoinCorrectness}, \ref{lem:unknown-access-transformer-correctness}, \ref{lem:access-transformer-correctness}, and \ref{lem:statement-transformer-correctness}:
\begin{restatable}[Analysis Correctness]{theorem}{MainTheorem}
	For all $v \in V$, we have:
	\begin{equation*}
		R^C(v) \subseteq \gamma(\widehat{R}(v))  
	\end{equation*}
\end{restatable}

\section{Loop Peeling and Unrolling}\label{sec:peelingunrolling}

    \begin{figure*}
    	\centering
	\begin{tikzpicture}[yscale=1]
		\node[circ] (peel0) {};
		\node[circ, right=of peel0, yshift=0cm, xshift=0cm] (peel1) {};
		\node[circ, right=of peel1, yshift=0cm, xshift=0cm] (peel2) {};
		\node[circ, right=of peel2, yshift=0cm, xshift=1cm] (peeln) {};
		\node[circ, right=of peeln, yshift=0cm, xshift=1cm] (unroll0) {};
		\node[circ, right=of unroll0, yshift=0cm, xshift=1cm] (unroll1) {};

		\draw[e] (peel0) node[above,yshift=1mm] {$\peel_0$} node[below, yshift=-3mm] {$\{0\}$} node[below,yshift=-3mm,xshift=-1.77cm] {Loop iterations:} node[below,yshift=3mm,xshift=-1.3cm] {Contexts:}  -- (peel1);
		\draw[e] (peel1) node[above,yshift=1mm] {$\peel_1$} node[below, yshift=-3mm] {$\{1\}$} -- (peel2);
		\draw[e, dotted] (peel2) node[above,yshift=1mm] {$\peel_2$} node[below, yshift=-3mm] {$\{2\}$} -- (peeln);
		\draw[e] (peeln) node[above,yshift=1mm] {$\peel_{15}$} node[below, yshift=-3mm] {$\{15\}$} node[below,yshift=-4mm,xshift=-1.25cm] {\dots} node[above,yshift=2mm,xshift=-1.25cm] {\dots} -- (unroll0);

		\draw[e] (unroll0) node[above,yshift=1mm] {$\unroll_0$} node[below, yshift=-3mm] {$\{16,18,\dots\}$} -- (unroll1);
		\draw[e] (unroll1) node[above,yshift=1mm] {$\unroll_1$} node[below, yshift=-3mm] {$\{17,19,\dots\}$} -- ($(unroll1)+(0.75, 0.0)$) -- ($(unroll1)+(0.75, 1)$) -| ($(unroll0)+(0.0, 0.55)$);
 
	\end{tikzpicture}
	\myvspace{-1mm}
	\caption{Peeling and unrolling contexts and their corresponding loop iterations.} %
	\label{fig:peeling-unrolling} %
	\myvspace{-1mm}
    \end{figure*}
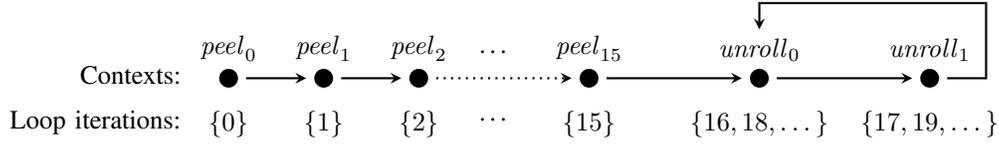

A common problem that cache analyses by abstract interpretation suffer from is the loss of precision due to joins at the entry of loops.
Indeed, the memory blocks loaded before a loop and within a loop usually differ. 
As a consequence the abstract cache states entering the loop and upon back edges from within the loop often have few, if any, memory blocks in common.
A sound analysis can thus not conclude any blocks to be cached at the beginning of the loop body.
One can avoid this issue by \emph{loop peeling}, where the analysis distinguishes the first few iterations of the loop from the rest of the loop and maintains separate analysis information for each of these iterations.
This allows the analysis to capture the ``warm-up effect'' commonly observed in loops iterating across arrays.
The example in Figure~\ref{fig:peeling-unrolling} shows a loop for which the first 16 loop iterations are peeled, which is the optimal amount of peeling for our example from Section~\ref{sec:illustrativeexample}.

Another problem that the basic analysis described in Section~\ref{sec:symbolic_analysis} suffers from is the lack of alignment information when establishing the alias relations between MCRs.
For example, one cannot tell whether $A[i]$ and $A[i+1]$ map to the same block if no information about the alignment of $A[i]$ is available. %
Indeed, it can happen that $A[i]$ and $A[i+1]$ are separated by a block boundary when $A[i] \bmod \BlockSize = \BlockSize - 1$.
The necessary alignment information can be obtained by \emph{unrolling loops}, i.e. distinguishing consecutive loop iterations from each other.
In the example in Figure~\ref{fig:peeling-unrolling} the loop is unrolled twice, distinguishing even from odd loop iterations.
In our example from Section~\ref{sec:illustrativeexample} we assumed a block size of $8$ bytes and array cells of size $4$ bytes.
Provided knowledge about the base address of the array $A$, with loop unrolling, the alignment of accesses to $A[i]$ is fully determined.\looseness=-1

\subsection{Context-Sensitive Analysis}

\newcommand{\Contexts}{\mathit{Ctxts}}
\newcommand{\Tags}{\mathit{Tags}}
\newcommand{\Ctx}{\mathit{ctx}}

Given peeling and unrolling depths $\MaxPeel \geq 0$ and $\MaxUnroll > 0$, we define the following set of tags: 
\begin{align*}
	\Tags := \{\peel_x \mid 0 \leq x < \MaxPeel\} ~\cup \\   \{\unroll_x \mid 0 \leq x < \MaxUnroll\}
\end{align*}
These correspond to the nodes in the graph in Figure~\ref{fig:peeling-unrolling}.
We then define contexts as functions that associate a tag with each loop variable, i.e.,  %
$\Contexts = \LoopVars \rightarrow \Tags$.
Then, $\peel_x$ means that the loop variable has value $x$, and $\unroll_x$ means that value of the loop variable is in $\{\MaxPeel + \MaxUnroll \cdot n + x \mid n \in \mathbb{N}\}$.

\newcommand{\CtxCaches}{\widehat{\mathit{SymCaches}}}
\newcommand{\CtxCache}{\widehat{\sigma}}
\newcommand{\ctxupdates}{\aupdates}
\newcommand{\ctxupdatea}{\aupdatea}

To avoid the precision loss at joins we lift our abstract domain to a context-sensitive domain~$\CtxCaches$ that associates a symbolic cache state with each context:
$$\CtxCaches = \Contexts \hookrightarrow \SymCaches$$
These abstract states are updated as follows %
 upon statements:
\begin{equation*}
\begin{multlined}
	\ctxupdates(\CtxCache, \Entry{i}) := \lambda \Ctx \in \Contexts. \\
		\begin{cases} 
			\bigsqcup_{t \in \Tags} \aupdates(\CtxCache(\Ctx[i \mapsto t]), \Entry{i})	& \text{if } \Ctx(i) = \peel_0\\
			\bot		& \text{otherwise}
		\end{cases}
\end{multlined}
\end{equation*}
Entering loop~$i$ corresponds to setting the loop variable~$i$ to zero.
Thus, independently, of the previous tag for $i$, the new tag for $i$ will be $\peel_0$.
The abstract value for this context is obtained by merging the values of all predecessor contexts, where $i$ may be arbitrary (first case).
Contexts in which the tag for $i$ is not $\peel_0$ are unreachable via entry edges (second case).\looseness=-1

To define the update upon back edges we first capture the structure of the graph in Figure~\ref{fig:peeling-unrolling} via its set of edges~$\mathcal{E}$:
\begin{align*}	
	\mathcal{E} :=  & \phantom{~}\{(\peel_x, \peel_{x+1}) \mid 0 \leq x < \MaxPeel-1\}\\
		&\cup \{(\peel_{\MaxPeel-1}, \unroll_0)\}\\
		&\cup \{(\unroll_x, \unroll_{x+1}) \mid 0 \leq x < \MaxUnroll-1 \}\\
		&\cup \{(\unroll_{\MaxUnroll-1}, \unroll_0)\}
\end{align*}
The set $\mathcal{E}$ captures how contexts evolve when taking back edges.
Based on $\mathcal{E}$ we define $\ctxupdates(\CtxCache, \Backedge{i})$:
\begin{equation*}
\begin{multlined}
	\ctxupdates(\CtxCache, \Backedge{i}) := \lambda \Ctx \in \Contexts.\hspace{9mm}\\
		\hspace{9mm}\bigsqcup_{\substack{\Ctx(i) = t'\\(t,t') \in \mathcal{E}}} \aupdates(\CtxCache(\Ctx[i \mapsto t]), \Backedge{i})	
\end{multlined}
\end{equation*}

Assume statements and memory accesses do not modify loop variables.
Thus, the update is simply applied pointwise %
 to each context.

\subsection{Refining Alias Relations using Context Information}

\newcommand{\evalmod}{\mathit{eval}_{\bmod}}
\newcommand{\exact}{\mathit{Exact}}
\newcommand{\knownmod}{\mathit{Mod}}
\newcommand{\unknown}{\mathit{Unknown}}

Contexts provide information about the values of loop variables, which can be used to deduce the alignment of MCRs.
To do so, we rely on an auxiliary function $\evalmod(e,\Ctx)$ that partially evaluates an MCR~$e$ in context $\Ctx$ obtaining one of the following results:\todo{mention this is in auxiliary material?}
\begin{itemize}
	\item $\exact(n)$, if the MCR is known to be exactly equal to $n$ in context $\Ctx$.
	\item $\knownmod(n, p)$, if the MCR is known to be equal to $n$ modulo~$p$ in context $\Ctx$.
	\item $\unknown$ if no such statement can be deduced.
\end{itemize}

We omit $\evalmod$ here for brevity; its definition is provided in the 
\ifconfversion
extended version~\cite{arXivVersion}.
\else
appendix.
\fi

Using $\evalmod$, we can refine the $\alias$ function and use the context to deduce alignment relations.
Given two MCRs $e_1$ and $e_2$, and a context $\Ctx$, we refine $\alias$ as follows:
\begin{equation*}
	\begin{multlined}
	\alias(e_1, e_2, \Ctx) :=\\ \begin{cases}
		\SameBlock
			& \text{if } n = e_1 - e_2 \in \mathbb{Z} \wedge
				     a_1 \sqsubseteq \knownmod(n_1, \BlockSize) \\ &~ \wedge~
				     a_2 \sqsubseteq \knownmod(n_2, \BlockSize) \wedge
				     n - n_1 + n_2= 0\\
		\SameSet
			& \text{if } n = e_1 - e_2 \in \mathbb{Z} \wedge
			             a_1 \sqsubseteq \knownmod(n_1, \BlockSize) \\ &~ \wedge~
				     a_2 \sqsubseteq \knownmod(n_2, \BlockSize) \\ &~ \wedge~
				     n - n_1 + n_2 \bmod \NbSets \cdot \BlockSize = 0\\
		\DifferentSet
			& \text{if } n = e_1 - e_2 \in \mathbb{Z} \wedge
			             a_1 \sqsubseteq \knownmod(n_1, \BlockSize) \\ &~ \wedge~
				     a_2 \sqsubseteq \knownmod(n_2, \BlockSize) \\ &~ \wedge~
				     n - n_1 + n_2 \bmod \NbSets \cdot \BlockSize \neq 0\\
		\alias(e_1, e_2)\hspace{-1cm}~
			&  \hspace{1cm}\text{otherwise}
		\end{cases}
	\end{multlined}
\end{equation*}
where $a_1 = \evalmod(e_1, \Ctx)$ and $a_2 = \evalmod(e_2, \Ctx)$, $\exact(k) \sqsubseteq \knownmod(n, m)$ if $k = n \bmod m$, and $\knownmod(n', m') \sqsubseteq \knownmod(n, m)$ if $m \vert m'$ and $n = n' \bmod m$. %

This refined alias function first looks at the difference $e_1 - e_2$ just like the non-refined version, except that the conditions to derive some relations are relaxed if the alignments ($a_1$ and $a_2$) of $e_1$ and $e_2$ are known.

In the first case, $n_1$ and $n_2$ are the offsets of $e_1$ and $e_2$ in their respective blocks.
Thus, one can deduce the address of the block that $e_1$ maps to ($e_1 - a_1$), and compare it to the address of the block that $e_2$ maps to ($e_2 - a_2$).
The equality of block addresses can be rewritten $n -n_1 + n_2 = 0$.
If the equality holds, then $e_1$ and $e_2$ map to the same block.

\todo{the second and third case should probably take $\knownmod(n_1, \BlockSize)$ and $\knownmod(n_2, \BlockSize)$!}

The second case is similar, but we check an equality on cache sets instead of blocks.
We thus consider alignments relative to sets, by evaluating $e_1$ and $e_2$ modulo $\NbSets \cdot \BlockSize$.
The equality is also checked modulo the same value because addresses that are $\NbSets \cdot \BlockSize$ apart map to the same set.

The third case is analogous, except we check for expressions mapping to different sets instead of the same one.
Finally, in cases were  $\evalmod$ fails to evaluate $e_1$ and $e_2$ precisely, we rely on the version of $\alias$ from Section~\ref{sec:abstrans} as a fallback.

\todo{WCET Analysis based on Implicit Path Enumeration}
\todo{maybe briefly discuss how the peeling and unrolling affects WCET analysis...?}

\section{Implementation}\label{sec:implementation}

We implemented the symbolic analysis in LLVMTA~\cite{Hahn2018,Hahn2022,Hahn2022b}, a WCET analysis tool based on the LLVM compiler infrastructure.
In particular, LLVMTA relies on LLVM to compile the program, which itself uses ScalarEvolution~\cite{ScalarEvolution,Absar2018} to perform optimizations.
It was thus convenient to reuse this framework and convert ScalarEvolution expressions to our own MCR representation upon which we added support for the shifting and substitution operations.
The main difficulty arising when converting ScalarEvolution expressions to MCRs is that ScalarEvolution (SCEV) expressions do not only contain integer constants but also LLVM values that belong to the LLVM intermediate representation (IR).
Consider an array $A$ that is allocated on the stack in a function $f$ and then passed down to another function $g$ accessing $A[i]$.
A SCEV expression for such an access would typically look like $\{\%A, +, 4\}_i$, where $\%A$ is a parameter of $f$.
We rely on debug information to determine the register containing the value of $\%A$, and then query a dedicated constant value analysis to get the register value.
This allows us to translate information available at the IR level down to the machine-code level at which our analysis is performed.\looseness=-1

Several tricks are implemented to make the analysis more efficient.
First, we rely on hash consing ({\url{https://en.wikipedia.org/wiki/Hash_consing}) of MCRs to reduce the memory footprint of the analysis: when building an MCR, we check if it was already build before, and return a pointer on the old MCR when possible.
In addition to saving memory, this allows us to cache and reuse the results of all operations involving MCRs. %

Another trick to speed up the analysis is to avoid representing a symbolic cache state $\SymCache \in \SymCaches$ as a single map of MCRs to ages.
Instead, a cache state is split into several maps, which we called ``virtual sets''.
We use one virtual set per physical cache set to store expressions that are known to map to this cache set.
An additional virtual set is used for expressions whose corresponding cache set is unknown.
When looking for ``same block'' MCRs (e.g. in $\aupdatea$), MCRs that map to a different virtual set than the accessed MCR can be excluded from the check, saving time.
Virtual sets can also be shared between abstract states.
Upon a memory access, if the set to which the accessed MCR is known, only the corresponding virtual set is modified.
The remaining virtual sets can thus be shared between the old and the new abstract state, saving memory and avoiding copies.

Regarding the values of $\MaxPeel$ and $\MaxUnroll$, it is not possible to choose fixed values that would work well for every benchmark due to the presence of nested loops. For example, it is possible to peel the first 256 iterations of a single loop, but doing so for each loop of a loop nest of depth 3 would lead to the creation of $256^3$ different contexts, blowing up the analysis complexity.
We thus introduce the notion of a \emph{peeling budget} in the analysis, which indicates the number of peeling contexts to create per loop nest.
This budget is first spent on the innermost loop, then on the second innermost loop if it is possible to fully peel the innermost one, and so on.
For example, consider a loop nest of depth 2, with loop bounds of 20 and 50 for the outer and inner loops, respectively.
A peeling budget of 200 would lead to fully peeling the inner loop, because the loop bound of the inner loop is less than the current budget.
Then the budget remaining for the outer loop would be 200/50, leading to a $\MaxPeel$ value of 4 for the outer loop.
We could introduce a similar notion for computing the $\MaxUnroll$ value associated to each loop.
Because this seemed unnecessary in many benchmarks, we chose to only unroll the innermost loop.

\section{Experimental Evaluation}\label{sec:experiments}

The aim of our experiments is to evaluate the following three aspects of our contributions:
\begin{enumerate}
	\item The gain in accuracy obtained by performing cache analysis over a symbolic CFG.
	\item Scalability when increasing the dataset sizes.
	\item Scalability in terms of the cache geometry.
\end{enumerate}

\todo{comparison with relational cache analysis}

First, we demonstrate the properties of our analysis on the illustrative example from Section~\ref{sec:illustrativeexample}.
Then, we present experiments designed to assess the accuracy gain due to the symbolic approach and its scalability.
All experiments are performed assuming a set-associative cache consisting of 8 cache sets,  8 cache ways, and cache lines of 64 bytes.
We qualitatively contrast our work with other related work in Section~\ref{sec:relatedwork}.\looseness=-1

In this evaluation we use the PolyBench~\cite{PolyBench} benchmarks.
PolyBench has the advantage of providing a parametric dataset size, i.e.\ one can adapt the sizes of the data structures the algorithms iterate over.
PolyBench provides 5 datasets size: \emph{mini}, \emph{small}, \emph{medium}, \emph{large}, and \emph{extra large}, which is convenient to assess the scalability of our approach.

\subsection{Behavior of the Symbolic Analysis on Illustrative Example}

\begin{figure*}
 	\begin{subfigure}[b]{0.49\textwidth}\centering
		\includegraphics[width=0.9\textwidth]{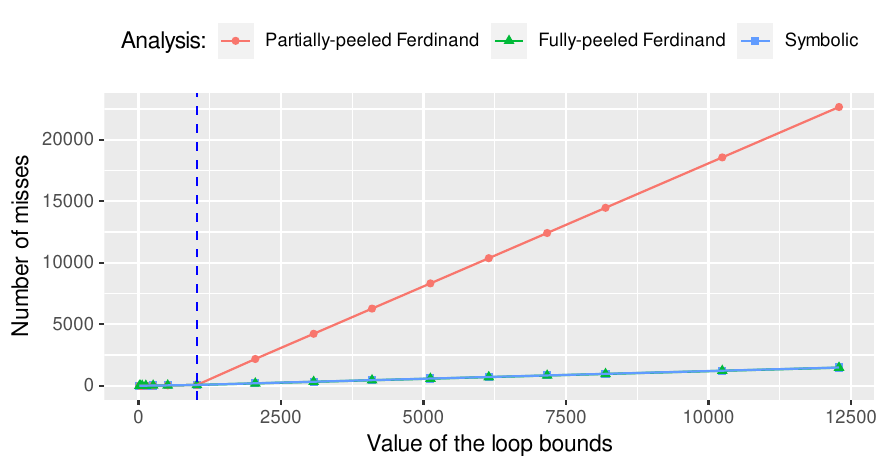}
		\caption{Accuracy comparison when increasing the dataset's size.}
		\label{fig:running_example_misses}
	\end{subfigure}\hfill
 	\begin{subfigure}[b]{0.49\textwidth}\centering
		\includegraphics[width=0.9\textwidth]{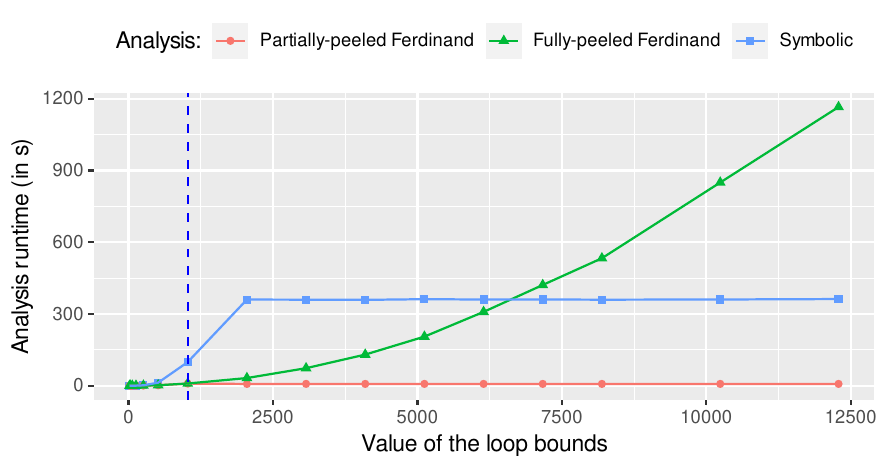}
		\caption{Analysis time comparison when increasing the dataset's size.}
		\label{fig:running_example_runtime}
	\end{subfigure}
	\myvspace{-1mm}
	\caption{Accuracy and analysis time comparison on the running example.}
	\myvspace{-1mm}
\end{figure*}

To verify that the symbolic analysis is behaving as expected, we analyze multiple variants of the program in Figure~\ref{fig:motivatingprogram} from Section~\ref{sec:illustrativeexample}.
In all experiments, we use an array of $12\cdot 1024=12288$ integers, but we vary the number of loop iterations in both loops between $4$ and $12288$, iterating back and forth across prefixes of the array.
We then compare the following analyses:
\begin{asparaitem}
	\item The symbolic analysis in optimal settings: we peel the exact number of iterations (1024) required to fill the cache, and we unroll enough iterations (128) to obtain perfect cache alignment information.
	\item Ferdinand's must analysis~\cite{Ferdinand1997,Ferdinand1999} under the same settings, i.e.\ using the same $\MaxPeel$ and $\MaxUnroll$ values.
	\item Ferdinand's analysis where both loops are fully peeled.
\end{asparaitem}
In each of these analyses we configure LLVMTA to compute a bound on the number of cache misses. 

We use Ferdinand's analysis as a baseline, as the symbolic analysis can be seen as a lifted version of Ferdinand's analysis to symbolic CFGs, and thus the observed differences can be directly attributed to operating symbolically.

Figure~\ref{fig:running_example_misses} shows the number of predicted misses when increasing the loop bounds.
As expected, for low values of the loop bounds all analyses fully peel the loops, and achieve the same perfect results: The first loop incurs one miss in every 16 iterations, as 16 consecutive integers of 4 bytes fit in a 64-byte cache line.
The second loop does not lead to any additional misses because the accessed data fits entirely in the cache.
Once the loop bounds are big enough to fill the cache, to the right of the dashed vertical line, the predicted number of misses  increases by 2 for every 16 loop iterations for both the symbolic analysis and Ferdinand's analysis if the loops are fully peeled.
This is due to additional misses at the end of the second loop, which accesses blocks that were evicted at the end of the first loop. 
Indeed, the results of the symbolic analysis and of Ferdinand's analysis under full peeling are exact.\looseness=-1

However, when the loop bounds exceed the number of peeled iterations, Ferdinand's analysis is unable to classify any access as a hit anymore.
As a consequence, the bound on the number of potential misses increases with every access: spatial locality is not exploited because the analysis does not know the offset of the accesses inside a cache line.

Figure~\ref{fig:running_example_runtime} shows the analysis runtime of the three analyses in terms of the loop bounds.
Once the loop bounds exceed the $\MaxPeel$ value, the analysis cost remains constant.
Conversely, when increasing the value of $\MaxPeel$ to match the loop bound, Ferdinand's analysis gets more and more expensive, quickly exceeding the cost of the symbolic analysis.\looseness=-1

\subsection{Accuracy of the Symbolic Analysis}

In order to evaluate the benefits of the symbolic analysis in more realistic cases, we analyze the PolyBench benchmarks (with the default dataset size \emph{large}), and compare its accuracy with Ferdinand's analysis. %
The cache configuration is fixed, but we vary the values of $\MaxPeel$ and $\MaxUnroll$.
Indeed, both analyses perform very differently in terms of running time and accuracy when varying the peeling and unrolling settings, and comparing the two for a fixed setting would thus be difficult.
So we set a runtime limit of one hour per benchmark and retain for each analysis the best achievable result within this time for each benchmark. 
Figure~\ref{fig:runtime_limited_misses} shows that in these conditions, the symbolic analysis  always outperforms Ferdinand's analysis.
The geometric mean of the ratios of the bounds computed by the symbolic and non-symbolic analysis across all benchmarks is $0.335$, significantly improving analysis accuracy.

\begin{figure}
	\centering
		\includegraphics[width=0.48\textwidth]{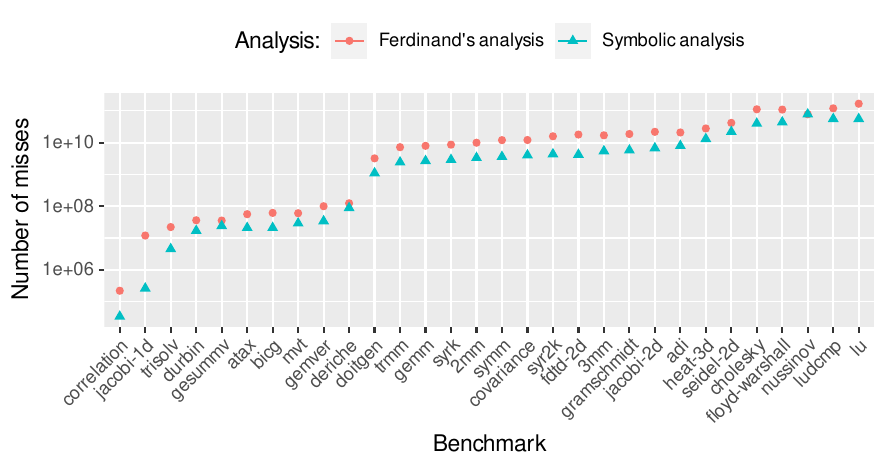}
		\myvspace{-1mm}
		\captionof{figure}{Accuracy comparison under a time constraint of 1 hour.}
		\label{fig:runtime_limited_misses}
		\myvspace{-2mm}
\end{figure}
\begin{figure}
	\centering
		\includegraphics[width=0.48\textwidth]{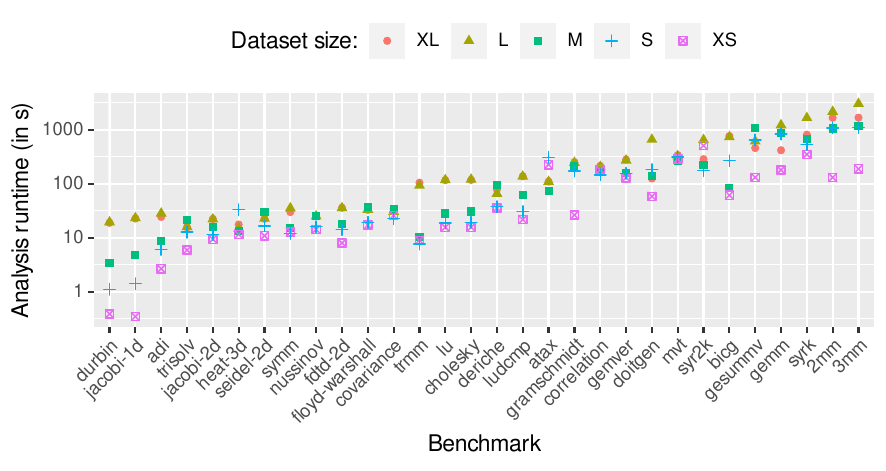}
		\myvspace{-1mm}
		\captionof{figure}{Analysis runtimes for increasing dataset sizes.}
		\label{fig:scalability_dataset}
		\myvspace{-2mm}
\end{figure}

\subsection{Scalability Evaluation}

We claim that the symbolic analysis runtime is largely independent of the number of loop iterations, as long as the number of loop iterations exceeds the number of peeled iterations.
To support this claim, we ran the analysis using the same cache configuration and peeling/unrolling settings ($\MaxPeel=1024$, $\MaxUnroll=128$) for all the dataset sizes available in PolyBench.
Figure~\ref{fig:scalability_dataset} shows the analysis runtime for each benchmark and dataset size.
Notice that the dataset size has a smaller impact on the analysis runtime than the benchmark itself, which suggests that the complexity of a benchmark's access patterns is more important than the number of accesses generated by the benchmark.
As expected, analysis times for the \emph{large} and \emph{extra large} datasets are usually very close to each other even though the number of memory accesses in the XL case is 6.25 times higher on the average.
For the smaller dataset sizes the loop bounds often do not reach the peeling settings, and thus the analysis cost still increases moving from XS to S, and sometimes also from S to M and~L.\looseness=-1

\todo{it might be nice to argue about the difference in the dynamic number of memory accesses between the different data sets sizes. that would allow to make a better case about scalability. maybe reason based on the simulation work that even just simulating (not static analysis) of the XL case would take very long...}

\subsection{Impact of the Cache Geometry}

To evaluate the impact of the cache geometry on the analysis runtime we designed two experiments.

In the first experiment, we investigate the impact of the associativity on the analysis runtime. 
We fix the cache line size to 64 bytes and the number of cache sets to 8, as in the previous experiments, and analyze associativities 8, 16, 32, and 64, corresponding to cache sizes of 4, 8, 16, and 32 KB, respectively.
We run the symbolic analysis on all benchmarks of PolyBench for the \emph{large} dataset.
To enable the analysis to exploit the increased cache size, we double the peeling budget each time we double the associativity.
Figure~\ref{fig:associativity} shows the geometric mean of the slowdowns relative to an analysis with associativity 8.
We observe a slowdown of 2.56, 10.7, and 70 at associativity 16, 32, and 64, respectively.

\begin{figure}
	\centering
	\begin{tikzpicture}
		\begin{axis}[
		  ybar,
		  bar width=15pt,
		  xlabel={Associativity ($\Associativity$)},
		  ylabel={Slowdown},
		  xtick=data,
		  xticklabels={$8$, $16$, $32$, $64$},
		  legend pos=north west,
		  grid=both,
		  ymode=log,
		  ymin=1,
		  height=4cm,
		  width=8cm
		]
	  
		\addplot[fill=blue] coordinates {
		  (1, 1)
		  (2, 2.56)
		  (3, 10.7)
		  (4, 70)
		};
		
		\legend{Associativity ($\Associativity$)}
	  
		\end{axis}
	\end{tikzpicture}
	  
\caption{Geometric mean of slowdowns relative to an analysis with associativity 8 across PolyBench for the \emph{large} dataset.}\label{fig:associativity}
\end{figure}
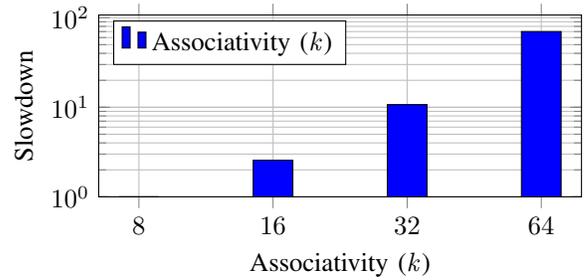

In the second experiment, we investigate the impact of the number of cache sets on the analysis runtime.
Thus, we fix the cache line size to 64 bytes and the associativity to 8, and perform analyses for 8, 16, 32, 64, and 128 cache sets, corresponding to cache sizes of 4, 8, 16, 32 and 64 KB, respectively.
Again, we double the peeling budget each time we double the number of cache lines.
Figure~\ref{fig:cachesets} shows the geometric mean of the slowdowns relative to an analysis with 8 cache sets.
We observe a slowdown of 2.07, 5.99, 23.8, and 125 at 16, 32, 64, and 128 cache sets, respectively. %
\looseness=-1

\begin{figure}
	\centering
	\begin{tikzpicture}
		\begin{axis}[
		  ybar,
		  bar width=12pt,
		  xlabel={Number of Sets ($\NbSets$)},
		  ylabel={Relative Slowdown},
		  xtick=data,
		  xticklabels={$8$, $16$, $32$, $64$, $128$}, %
		  legend pos=north west,
		  grid=both, %
		  ymode=log,
		  ymin=1,
		  height=5cm, %
		  width=8cm
		]
	  
		\addplot[fill=red] coordinates {
		  (1, 1)
		  (2, 2.07)
		  (3, 5.99)
		  (4, 23.8)
		  (5, 125)
		};
		
		\legend{Number of Sets ($\NbSets$)}
	  
		\end{axis}
	  \end{tikzpicture}
	  
\caption{Geometric mean of slowdowns relative to an analysis with 8 cache sets across PolyBench for the \emph{large} dataset.}\label{fig:cachesets}
\end{figure}
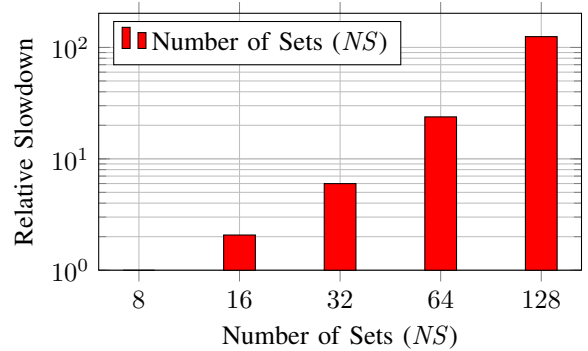

In both experiments, we observe that the analysis runtime increases superlinearly with the cache size.
Indeed, there are two effects at play here that are each individually expected to induce a linear slowdown: (i) the peeling budget is proportional to the cache size and thus the number of contexts increases linearly, and (ii) the abstract cache states grow linearly in the cache size. 
The effect of (ii) on the analysis runtimes is less pronounced when increasing the number of cache sets than when increasing the associativity due to the use of virtual sets, and we observe smaller slowdowns there.

\section{Related Work}\label{sec:relatedwork}

Static cache analysis has received considerable attention in the context of WCET analysis.
In the following, we focus on work targeted at data cache analysis.
For a broader review of the literature consider the survey paper by Lv et al.~\cite{Lv2016}.

At a high level, work on static cache analysis can be partitioned into classifying and bounding analyses:
\begin{asparaitem}
	\item \emph{Classifying analyses}~\cite{Ferdinand1997,Ferdinand1999,Sen07,Grund2009,Grund2010,Grund2010b,Chattopadhyay2013,Touzeau2017,Touzeau2019,Brandner2020,Brandner2022} classify individual accesses in the program as hits or misses. Ferdinand's may and must analysis and our symbolic analysis fall into this class.\looseness=-1
	\item \emph{Bounding analyses}~\cite{Kim1996,White1997,Ferdinand1999,Huynh11,Guan13,Guan14,Reineke18,Sotin21} compute bounds on the number of misses that occur in a program fragment or in a subset of the program's accesses. %
\end{asparaitem}

Let us first discuss related classifying analyses. %
We have already extensively discussed Ferdinand's LRU must analysis~\cite{Ferdinand1997,Ferdinand1999} throughout the paper.
It relies on a plain CFG abstraction, and precise analysis results for data caches are only possible if loops are fully unrolled. %

Sen and Srikant~\cite{Sen07} build upon LRU must analysis and make two contributions:
(i)~They introduce a new domain to analyze the set of memory addresses associated with a static memory reference called \emph{circular linear progressions}. %
(ii)~They introduce a new approach to context-sensitive analysis in which a loop is partitioned into $n$ same-length regions that are further split into two parts. The first part is analyzed in ``expansion mode'', meaning that it is fully virtually unrolled, distinguishing all individual iterations, while the second part is analyzed in ``summary mode''.
To achieve accurate results, the approach requires an unrolling value that is proportional to the number of loop iterations, similarly to Ferdinand's analysis.

Hahn and Grund~\cite{Hahn2011,Hahn12} introduce \emph{relational cache analysis}, which tracks relations between memory accesses in the program following the lattice in Figure~\ref{fig:alias-relation} similarly to our analysis. 
Wegener~\cite{Wegener2012} proposes to judiciously apply loop peeling and unrolling to relational cache analysis.
Their work is able to detect the exploitation of spatial and temporal locality within a given loop iteration (or within a sequence of loop iterations in case of unrolling).
The fundamental limitation of~\cite{Hahn2011,Hahn12,Wegener2012} that our approach overcomes, is that their analysis never tracks more than a single symbol for each static memory reference (per unrolled iteration of the loop) in the program, whereas our analysis may dynamically generate an unbounded number of symbols for the same static reference due to the shifting operation upon loop back edges. 
As a consequence, in our example program, the temporal locality in the second loop would be entirely missed by relational cache analysis.
The other major difference lies in our use of LLVM's ScalarEvolution framework to determine access expressions and loop bounds.

Let us now turn to bounding analyses.
Kim et al.~\cite{Kim1996} determine a bound on the number of memory blocks accessed in a program. 
If at most $m$ distinct blocks are accessed, and these fully fit into the cache, then at most $m$ misses may occur.
Such a cache persistence~\cite{Reineke18} argument only works in cases where the amount of accessed data is smaller than the cache itself, which is often not the case, e.g. in our illustrative example and in the entire PolyBench suite for larger dataset sizes.\looseness=-1

Huynh et al.~\cite{Huynh11} present a persistence analysis that takes a different perspective, separately considering each memory block accessed in the program.
For each such block, the analysis determines whether it is persistent, i.e., whether accesses to that block can result in more than one miss.
This persistence classification is furthermore performed at different spatial and temporal scopes, e.g. distinguishing different intervals of loop iterations.
As a result the analysis may be highly accurate. 
However the analysis complexity is at least linear in \emph{both} the number of distinct memory blocks accessed by the program and the dynamic number of accesses performed ($>10^{11}$ for several PolyBench benchmarks for the XL dataset), whereas our analysis is independent of both of these. %

The approach of Sotin et al.~\cite{Sotin21} consists in encoding the program semantics and the cache replacement policy in a formula whose integral solutions correspond to cache misses, and to discharge this counting problem to an external solver~\cite{Barvinok93}.
The approach is however limited to counting misses associated to a single static memory reference inside a loop.
Ad hoc extensions handling non-linear accesses, several accesses in the same loop, and analyzing nested loops are suggested, but it is not clear whether these approaches can be combined together to handle larger classes of programs.

Finally, there is a long and rich history of analytical cache models~\cite{Ghosh1997,Ghosh1999,Chatterjee2001,Vera2002,Vera2004,Cascaval2003,Beyls2005,Bao2018,Gysi2019,Morelli22} that determine the exact number of misses generated by loop nests.
A common limitation of this line of work is that it cannot handle programs with input-dependent branches or memory accesses.\looseness=-1

\section{Conclusions and Future Work}\label{sec:conclusions}

We have introduced \emph{symbolic data cache analysis} a novel analysis that systematically exploits a richer program abstraction than prior work, namely \emph{symbolic control-flow graphs}, which can be obtained from LLVM's ScalarEvolution analysis.
The experimental evaluation demonstrates that this new analysis outperforms classical LRU must analysis both in terms of accuracy and analysis runtime.

As a proof of concept, we have lifted the classical LRU must analysis to the symbolic level.
Other existing analyses operating on plain CFGs could similarly be made symbolic, e.g. persistence analyses or classifying analyses for various replacement policies.
It would also be interesting to investigate whether exact cache analysis on symbolic CFGs is possible along the lines of recent exact cache analyses on plain CFGs.

Another direction for future work is to apply the idea of symbolic cache analysis to even richer program abstractions, e.g. modeling operations on heap data structures.

\section*{Acknowledgments}
This project has received funding from the European Research Council under the EU's Horizon 2020 research and innovation programme (grant agreement No. 101020415).

\bibliographystyle{IEEEtran}
\balance
\bibliography{symbolic_data_cache_analysis.bib}

\ifconfversion
\else

\newpage

\appendices
\section{Proofs}

\JoinCorrectness*
\begin{proof}
	Let $\SymCache_1, \SymCache_2 \in \SymCaches$, and $(\pstate, \cstate) \in \gamma(\SymCache_1) \cup \gamma(\SymCache_2)$.
	We assume without loss of generality that $(\pstate, \cstate) \in \gamma(\SymCache_1)$.
	By definition of the concretization function, we have:
	\begin{equation*}
		\forall e \in \dom(\SymCache_1): \cstate(\block(\eval{e}{\pstate})) \leq \SymCache_1(e) \\
	\end{equation*}
	Thus, for any $e \in \dom(\SymCache_1) \cap \dom(\SymCache_2)$, we have:
	\begin{align*}
		\cstate(\block(\eval{e}{\pstate}))
		& \leq \SymCache_1(e) \\
		& \leq \max(\SymCache_1(e), \SymCache_2(e))\\
		& \leq (\SymCache_1 \sqcup \SymCache_2)(e)
	\end{align*}
	Thus, $(\pstate, \cstate) \in \gamma(\SymCache_1 \sqcup \SymCache_2)$.
\end{proof}

\UnknownAccessTransformerCorrectness*
\begin{proof}
	Let $\SymCache \in \SymCaches$.
	Let $(\pstate', \cstate') \in \update(\gamma(\SymCache), \unknownAccess)$.

	We will show that $(\pstate', \cstate') \in \gamma(\aupdatex(\SymCache))$.

	Let $(\pstate, \cstate) \in \gamma(\SymCache)$ and $b \in \blocks$ such that $(\pstate', \cstate') = \update((\pstate, \cstate), b)$.
	We know that $\forall e \in \dom(\SymCache) : \cstate(\block(\eval{e}{\pstate})) \leq \SymCache(e)$, and we want to prove that $\forall e \in \dom(\SymCache'). \cstate'(\block(\eval{e}{\pstate'})) \leq \SymCache'$, where $\SymCache' = \aupdatex(\SymCache)$.

	Let $e \in \dom(\SymCache') = \dom(\SymCache)$.
	Expanding all definitions, we have $\cstate'(\block(\eval{e}{\pstate'})) \leq \cstate(\block(\eval{e}{\pstate})) + 1 \leq \SymCache(e) + 1 \leq \SymCache'(e)$.

	Thus $(\pstate', \cstate') \in \gamma(\SymCache')$, finishing the proof.
\end{proof}

\AccessTransformerCorrectness*
\begin{proof}
	Let $\SymCache \in \SymCaches$, $e \in \MCR{\LoopVars}$ and $(\pstate, \cstate) \in \gamma(\SymCache)$.
	We will show that \\$(\pstate, \updatelru(\cstate, \block(\eval{e}{\pstate}))) \in \gamma(\aupdatea(\SymCache, e))$.
	To ease reading, we introduce the following notations:
	\begin{itemize}
		\item $b = \block(\eval{e}{\pstate})$, the block $e$ maps to.
		\item $\SymCache' = \aupdatea(\SymCache, e)$, the successor of $\SymCache$ after the update.
		\item $\cstate' = \updatelru(\cstate, b)$, the successor of $\cstate$.
	\end{itemize}

	We then want to prove that: $(\pstate, \cstate') \in \gamma(\SymCache')$, i.e.\
	$\forall e': \cstate'(\block(\eval{e'}{\pstate})) \leq \SymCache'(e')$.

	Let $e' \in \dom(\SymCache) \cup \{e\}$.
	Noting $b' = \block(\eval{e'}{\pstate})$, we want to show that $\cstate'(b') \leq \SymCache'(e')$.
	This is done by reasoning by case distinction, looking at how $\SymCache'(e')$ is obtained from $\SymCache(e')$.

	\begin{itemize}
		\item Assume $\alias(e, e') \sqsubseteq \SameBlock$:
			By correctness of $\alias$, we have: $b = b'$, which by definition of $\updatelru$ leads to $\cstate'(b') = 0$.
			We thus have $\cstate'(b') \leq \SymCache'(e')$.
			This case covers in particular the condition $e = e'$.
			In the remaining cases, we will thus assume that $e \neq e'$, and thus that $e' \in \dom(\SymCache)$.
		\item Suppose $\alias(e, e') \sqsubseteq \SameBlockWhenSameSet$:
			Again, by correctness of $\alias$, we have either $b = b'$ or $\set(b) \neq \set(b')$.
			If $b = b'$, we are back to the previous case and $\cstate'(b') = 0 \leq \SymCache'(e')$.
			Otherwise, $\set(b) \neq \set(b')$, and by definition of $\updatelru$ we obtain $\cstate'(b') = \cstate(b')$.
			However, we have $\cstate(b') \leq \SymCache(e')$ because $(\pstate, \cstate) \in \gamma(\SymCache)$ and $e' \in \dom(\SymCache)$.
			In addition, by case hypothesis, and definition of $\aupdatea$, we have: $\SymCache'(e') = \SymCache(e')$.
			Combining these inequalities together we obtain:
			\begin{equation*}
				\cstate'(b') = \cstate(b') \leq \SymCache(e') = \SymCache'(e')
			\end{equation*}
		\item Otherwise, assume $\SymCache(e) \leq \SymCache(e')$:
			By $\aupdatea$ and $\gamma$, we get $\SymCache'(e') = \SymCache(e') \geq \cstate(b')$.
			We then do one additional case distinction:
			\begin{itemize}
				\item If $\cstate(b) \leq \cstate(b')$, by $\updatelru$ we obtain:
				\begin{equation*}
					\cstate'(b') = \cstate(b') \leq \SymCache(e') = \SymCache'(e')
				\end{equation*}
				\item Else $\cstate(b') < \cstate(b)$. We thus have:
				\begin{equation*}
						\cstate'(b') \leq \cstate(b') + 1 \leq \cstate(b) \leq \SymCache(e) \leq 						\SymCache(e') = \SymCache'(e')
				\end{equation*}
			\end{itemize}
			Both subcases thus lead to $\cstate'(b') \leq \SymCache'(e')$ as required.
		\item Otherwise, suppose $\SymCache(e') + 1 < \Associativity$:
			By definition of $\aupdatea$, we have $\SymCache'(e') = \SymCache(e') + 1$.
			Considering all cases in $\updatelru$, observe that $\cstate'(b') \leq \cstate(b')+1$, and so
			\[\cstate'(b') \leq \cstate(b')+1 \leq \SymCache(e')+1 = \SymCache'(e')\]
						
		\item Finally, assume none of the conditions above apply:
			By definition $\aupdatea$, we have $\SymCache'(e') = \infty$ and so trivially $\cstate'(b') \leq \SymCache'(e')$.
	\end{itemize}
	This finishes the proof. In all cases, we have $\cstate'(b') \leq \SymCache'(e')$ proving that $(\pstate, \cstate') \in \gamma(\SymCache')$, and thus:
	\begin{equation*}
		\update(\gamma(\SymCache), e) \subseteq \gamma(\aupdatea(\SymCache, e)) 
	\end{equation*}%
	\todo{This proof relies on the correctness of the alias relation, which remains to do}%
\end{proof}

\StatementTransformerCorrectness*
\begin{proof}
	Let $\SymCache \in \SymCaches$, $s \in \Statements$ and $(\pstate', \cstate') \in \update(\gamma(\SymCache), s))$.
	We will show that $(\pstate', \cstate') \in \gamma(\aupdates(\SymCache, s))$.
	In the remaining, we will note $\SymCache' = \aupdate(\SymCache, s)$.

	By definition of $\update$, $\pstate' \neq \unreachable$ and there exists $(\pstate, \cstate) \in \gamma(\SymCache)$ such that $(\pstate', \cstate') = \update((\pstate, \cstate), s)$.
	From this, we deduce $\cstate' = \cstate$. and $\pstate' = \updates(\pstate, s)$.

	We want to show that for any $e' \in \dom(\SymCache')$, $\cstate(\block(\eval{e'}{\pstate'})) \leq \SymCache'(e')$.

	Let $e' \in \dom(\SymCache')$.
	We proceed by case distinction on the value of $s$:

	\begin{itemize}
		\item If $s = \Entry{i}$ for some $i$:
			Because $e' \in \dom(\SymCache')$, we have $i \notin e'$ and thus $\eval{e'}{\pstate'} = \eval{e'}{\pstate[i \mapsto 0]} = \eval{e'}{\pstate}$.
			On the other side, we have: $\SymCache'(e) = \SymCache(e)$.
			Inserting these equalities in the definition of $(\pstate, \cstate) \in \gamma(\SymCache)$, we obtain the desired inequality: $\cstate(\block(\eval{e'}{\pstate'})) \leq \SymCache'(e')$.

		\item If $s = \Backedge{i}$ for some $i$:
			By definition of $\aupdates$, there is an $e$ such that $e' = \shift{e}{i}$ and $\SymCache'(e') = \SymCache(e)$.
			We thus have:
			\begin{align*}
				\cstate(\block(\eval{e'}{\pstate'}))
				& = \cstate(\block(\eval{\shift{e}{i}}{\pstate[i \mapsto \pstate(i) + 1]}))\\
				& = \cstate(\block(\eval{e}{\pstate})) \\
				& \leq \SymCache(e) \\
				& \leq \SymCache'(e')
			\end{align*}
		\item If $s = \Assume{i}{\mathit{expr}}$ for some $i$ and $\mathit{expr} \in MCR(\LoopVars \setminus \{i\})$:
			By defintion of $\aupdates$, we know there is an $e$ such that $e' = \subst{e}{i}{\mathit{expr}} \neq \mathit{fail}$ and $\SymCache'(e') = \SymCache(e)$.
			We already deduced that $\pstate' \neq \unreachable$, which now implies that $\pstate' = \pstate$ and $\pstate(i) = \eval{\mathit{expr}}{\pstate}$
			We thus have:
			\begin{align*}
				\cstate(\block(\eval{e'}{\pstate'}))
				& = \cstate(\block(\eval{\subst{e}{i}{\mathit{expr}}}{\pstate}))\\
				& = \cstate(\block(\eval{e}{\pstate[i \mapsto \eval{\mathit{expr}}{\pstate}]})) \\
				& = \cstate(\block(\eval{e}{\pstate[i \mapsto \pstate(i)]})) \\
				& = \cstate(\block(\eval{e}{\pstate})) \\
				& \leq \SymCache(e) \\
				& \leq \SymCache'(e')
			\end{align*}
	\end{itemize}
	In every case, it thus holds that $\cstate(\block(\eval{e'}{\pstate'})) \leq \SymCache'(e')$, proving that $(\pstate', \cstate') \in \gamma(\SymCache')$ as desired.
\end{proof}

\begin{lemma}[Concrete Domain Completeness]
	\label{thm:concrete-domain-completeness}
	The set $D = \powerset{(\LoopVars \rightarrow \mathbb{N}) \times (\mathcal{B} \rightarrow \mathbb{N})}$, where $\mathcal{B}$ is the set of memory blocks accessed by the program, is a complete lattice.
\end{lemma}
\begin{proof}
	$D$ being a powerset, $\bigcup S$ and $\bigcap S$ are obvious least upper and greatest lower bound for any subset $S$ of $D$.
\end{proof}

\begin{lemma}[Abstract Domain Completeness]
	\label{thm:abstract-domain-completeness}
	The set $\SymCaches = \MCR{\LoopVars} \hookrightarrow \{0, \dots, \Associativity-1, \infty\}$ is a complete lattice.
\end{lemma}
\begin{proof}
	Let $S \subseteq \SymCaches$.
	Consider $\bigsqcup S = \lambda e \in \bigcap_{\SymCache \in S} \dom(\SymCache). \max(\{\SymCache(e), \SymCache \in S\})$.
	$\bigsqcup S$ is well defined even for an infinite subset $S$ because $\{0, \dots, \Associativity-1, \infty\}$ being finite, $\{\SymCache(e), \SymCache \in S\}$ always admit a maximum.
	$\bigsqcup S$ belongs to $\SymCaches$ and the upper-bound property $\forall \SymCache \in S: \SymCache \sqsubseteq \bigsqcup S$ is obvious.
	One can show in a similar way that $\bigsqcap S = \lambda e. \min(\{\SymCache(e), \SymCache \in S\})$ is a lower bound in $\SymCaches$ for any subset $S$ of $\SymCaches$, making it a complete lattice.
\end{proof}

\MainTheorem*
\begin{proof}
	The semantics $R^C$ and $\widehat{R}$ can be rewritten as the least fixpoints of the following functions:
	\begin{align*}
		F(R) & =
			\lambda v. R_0 \cup \bigcup_{(u, d, v) \in E} \update(R(u), d)\\
		\widehat{F}(\widehat{R}) & = 
			\lambda v. \widehat{R_0}(v) \sqcup \bigsqcup_{(u, d, v) \in E} \aupdate(\widehat{R}(u), d)\\
		\text{where}\\
		R_0(v) & = \begin{cases}
			\{(\lambda i. 0, \cstate) \mid \cstate \in \cstates\} & \text{if } v = v_0\\
			\emptyset & \text{otherwise}
		\end{cases}\\
		\widehat{R_0}(v_0) & = \emptyset\\
	\end{align*}
	This is well-defined because the Knaster-Tarski fixpoint theorem guarantees the existence of these fixpoints.
	Indeed, $F$ and $\widehat{F}$ are monotone by construction, and Lemmas~\ref{thm:concrete-domain-completeness} and~\ref{thm:abstract-domain-completeness} ensures that our concrete and abstract domains are complete lattice.
	In addition, the Knaster-Tarski theorem ensures that $R^C = \mathit{lfp}(F) = \bigcap \mathit{Red}(F) = \bigcap \{R \mid R \supseteq F(R)\}$ (the same relation holds for $\widehat{R}$ and $\widehat{F}$ but we will not need it).

	From Lemmas~\ref{lem:unknown-access-transformer-correctness} and \ref{lem:access-transformer-correctness} and \ref{lem:statement-transformer-correctness} about the correctness of statement and access transformers, one can trivially derive that for any~$\SymCache$:
	\begin{equation*}
		\update(\gamma(\SymCache, d) \subseteq \gamma(\aupdate(\SymCache, d))
	\end{equation*}

	We thus have:
	\begin{align*}
		F(\gamma(\widehat{R}))
		& = \begin{multlined}[t]
			\lambda v. R_0 \cup \\
			\bigcup_{(u, d, v) \in E} \update(\gamma(\widehat{R}(u)), d)
		\end{multlined}\\
		& \subseteq \lambda v. R_0 \cup \bigcup_{(u, d, v) \in E} \gamma(\aupdate(\widehat{R}(u), d))\\
		& \subseteq \lambda v. R_0 \cup \gamma\left( \bigsqcup_{(u, d, v) \in E} \aupdate(\widehat{R}(u), d)\right)\\
		& \subseteq \lambda v. \gamma(\widehat{R_0}) \cup \gamma\left(\bigsqcup_{(u, d, v) \in E} \aupdate(\widehat{R}(u), d)\right)\\
		& \subseteq \lambda v. \gamma\left(\widehat{R_0} \sqcup \bigsqcup_{(u, d, v) \in E} \aupdate(\widehat{R}(u), d)\right)\\
		& \subseteq \gamma(\widehat{F}(\widehat{R}))
	\end{align*}

	Finally, we get the following inequalities:
	\begin{align*}
		\widehat{R} = \mathit{lfp}(\widehat{F})
		& \Rightarrow \widehat{F}(\widehat{R}) = \widehat{R}\\
		& \Rightarrow \gamma(\widehat{F}(\widehat{R})) = \gamma(\widehat{R}) \\
		& \Rightarrow F(\gamma(\widehat{R})) \subseteq \gamma(\widehat{R})\\
		& \Rightarrow \gamma(\widehat{R}) \in \mathit{Red}(F)\\
		& \Rightarrow \gamma(\widehat{R}) \supseteq \mathit{lfp}(F) = R^C
	\end{align*}
\end{proof}

\onecolumn

\section{Definition of $\evalmod$}
\newcommand{\evalmodaux}{\mathit{evalAux}_{\bmod}}

This section provides a formal definition of the $\evalmod$ function used in the submission.
The role of $\evalmod$ is to evaluate an expression in a context that provides partial  information about the current loop iteration.
Thus, the return value of $\evalmod$ needs to represent partially known values, which is done using the following enumeration:
\begin{itemize}
	\item $\exact(n)$ when the evaluated expression is known to be exactly $n$ in the given context.
	\item $\knownmod(n, k)$ when only the residue modulo $k$ is known to be $n$.
	\item $\unknown$ when no information can be derived in the given context.
\end{itemize}

We overload arithmetic operations (with $\mathit{bop} \in \{+, -, \times\}$) to operate on this partial knowledge as follows:
\begin{equation*}
	a_1~\mathit{bop}~a_2 =
		\begin{cases}
			\exact(n_1~\mathit{bop}~n_2) & \text{if } a_1 = \exact(n_1) \wedge a_2 = \exact(n_2) \\
			\knownmod((n_1~\mathit{bop}~n_2) \bmod k_2, k_2) & \text{if } a_1 = \exact(n_1) \wedge a_2 = \knownmod(n_2, k_2) \\
			\knownmod((n_1~\mathit{bop}~n_2) \bmod k_1, k_1) & \text{if } a_1 = \knownmod(n_1, k_1) \wedge a_2 = \exact(n_2) \\
			\knownmod((n_1~\mathit{bop}~n_2) \bmod \gcd(k_1, k_2), \gcd(k_1, k_2)) & \text{if } a_1 = \knownmod(n_1, k_1) \wedge a_2 = \knownmod(n_2, k_2) \wedge \\
			& ~\gcd(k_1, k_2) > 1\\
			\unknown & \text{otherwise}\\
		\end{cases}\\
\end{equation*}

We then define an auxiliary function $\evalmodaux$ as follows:
\begin{equation*}
	\begin{split}
		\evalmodaux(n,\Ctx,S) &:= \exact(n) \\
		\evalmodaux(e_1~\mathit{bop}~e_2,\Ctx,S) &:= \evalmodaux(e_1,\Ctx,S)~\mathit{bop}~\evalmodaux(e_2,\Ctx,S)\\
		\evalmodaux(\{e_1, +, e_2\}_i,\Ctx,S) &:=
			\begin{cases}
				\unknown & \text{if } i \in S\\
				\evalmodaux(e_1,\Ctx,S) + &
					\text{if } i \notin S \wedge \Ctx(i) = \peel_n \\
				~\evalmodaux(e_2,\Ctx,S \cup \{i\}) \times \exact(n) & \\
				\evalmodaux(e_1,\Ctx,S) + & \text{if } i \notin S \wedge \Ctx(i) = \unroll_n \\
				~\evalmodaux(e_2,\Ctx,S \cup \{i\}) \times &\\
				~\knownmod(\MaxPeel + n, \MaxUnroll) & \\
			\end{cases}
	\end{split}
\end{equation*}

The function $\evalmod$ is then defined as $\evalmod(e, \Ctx) = \evalmodaux(e, \Ctx, \{\})$.
The auxiliary function $\evalmodaux$ is recursive, and evaluates an expression by first evaluating its subexpressions in a bottom-up manner.
The set $S$, initially empty, is used to track down the set of induction variables already met along the evaluation path in the top-down direction.
Its role is to detect the evaluation of expressions of degree greater than $2$ for a given induction variable, and return $\unknown$ in this case.
Indeed, the provided formula would be invalid for such expressions when the loop induction variable is not exactly known.
Consider $\{1, +, \{2, +, 1\}_x\}_x$ as an example of such an expression, which represents $\frac{(x+1)(x+2)}{2}$, and assume $x$ is only known to be equal to $1$ modulo $2$.
$x=1$ would lead to $\frac{(x+1)(x+2)}{2} = 3$, but $x=3$ would lead to $\frac{(x+1)(x+2)}{2} = 10$, which has a different residue modulo $2$.
Note that the tracking of loop-induction variables to avoid this case is simple but incomplete: it is possible that $\evalmodaux$ returns $\unknown$ in cases in which it would be possible to extract better information.
However, in practice, most expressions are linear and thus $\evalmodaux$ almost always succeeds.
\fi

\end{document}